\documentclass[11pt, conference]{IEEEtran}
\IEEEoverridecommandlockouts
\renewcommand\IEEEkeywordsname{Keywords}

\usepackage{cite}
\usepackage{amsmath,amssymb,amsfonts}
\usepackage{algorithmic}
\usepackage{graphicx}
\usepackage{textcomp}
\usepackage{xcolor}
\usepackage{afterpage}

\usepackage{comment}
\usepackage{graphicx} 
\usepackage[colorinlistoftodos]{todonotes}
\usepackage[utf8]{inputenc}
\usepackage{multirow}
\usepackage{multicol}
\usepackage{slashbox}

\usepackage{authblk}
\usepackage{float} 
\usepackage{balance}
\usepackage{fancyhdr}
\usepackage{array}
\usepackage{wrapfig}
\usepackage{lscape}
\usepackage[figuresright]{rotating}
\ifCLASSOPTIONcompsoc
 \usepackage[caption=false,font=normalsize,labelfont=sf,textfont=sf]{subfig}
\else
 \usepackage[caption=false,font=footnotesize]{subfig}
\fi

\def\BibTeX{{\rm B\kern-.05em{\sc i\kern-.025em b}\kern-.08em
    T\kern-.1667em\lower.7ex\hbox{E}\kern-.125emX}}

\usepackage[english]{babel}
\usepackage[utf8]{inputenc}
\usepackage{fancyhdr}

\begin{document}

\title{Digital Twin for Secure Semiconductor Lifecycle Management: Prospects and Applications
}

\author{Hasan Al Shaikh}
\author{Mohammad Bin Monjil}
\author{Shigang Chen}
\author{Navid Asadizanjani}
\author{Farimah Farahmandi}
\author{\\Mark Tehranipoor}
\author{Fahim Rahman}
\affil{Email: \{hasanalshaikh,monjil.m\}@ufl.edu, Department of ECE, University of Florida \\
Email: sgchen@cise.ufl.edu, Department of CISE, University of Florida\\
Email: \{farimah,tehranipoor, fahimrahman, \}@ece.ufl.edu, Department of ECE, University of Florida
}

\maketitle
\thispagestyle{fancy}

\begin{abstract}
The expansive globalization of the semiconductor supply chain has introduced numerous untrusted entities into different stages of a device’s lifecycle, enabling them to compromise its security. 
To make matters worse, the increasing complexity in the design as well as aggressive time-to-market requirements of the newer generation of integrated circuits can lead either designers to unintentionally introduce security vulnerabilities or verification engineers to fail in detecting them earlier in design lifecycle, often due to the limitation of traditional verification and testing methodologies. 
These overlooked or undetected vulnerabilities can be exploited by malicious entities in subsequent stages of the lifecycle through an ever-widening variety of hardware attacks. 
The ability to ascertain the provenance of these vulnerabilities, after they have been unearthed at a later stage, becomes a pressing issue when the security assurance across the whole lifecycle is required to be ensured and generationally improved to thwart emerging attacks.

We posit that if there is a malicious or unintentional breach of security policies of a device, it will be reflected in the form of anomalies in the data collected through traditional design, verification, validation, and testing activities throughout the lifecycle.
With that, a digital simulacrum of a device's lifecycle, called a digital twin (DT), can be formed by the data gathered from different stages to secure the lifecycle of the device. 
The DT can analyze the collected data through its constituent AI and data analytics algorithms to trace the origin of a detected hardware attack or vulnerability to the associated stage of the lifecycle.
We refer to this functionality of the DT as \textit{Backward Trust Analysis}. 
We also introduce the notion of \textit{Forward Trust Analysis} which refers to the scalability and adaptability of the DT to unforeseen threats as they emerge.

In this paper, we put forward a realization of intertwined relationships of security vulnerabilities with data available from the silicon lifecycle and formulate different components of an AI driven DT framework. The proposed DT framework leverages these relationships to achieve aforementioned security objectives through causality analysis, and thus accomplish end-to-end security-aware management of the entire semiconductor lifecycle.
We put a perspective on how the limitations of existing ad-hoc-style security solutions can be overcome by the data oriented analysis that underpins our approach.

With several threat and attack scenarios, we demonstrate how advanced modeling techniques can perform relational learning to identify such attacks.
Finally, we provide potential future research avenues and challenges for realization of the digital twin framework to enable secure semiconductor lifecycle management.


\end{abstract}

\begin{figure*}[t]
\centering
\includegraphics[width=0.9\textwidth]{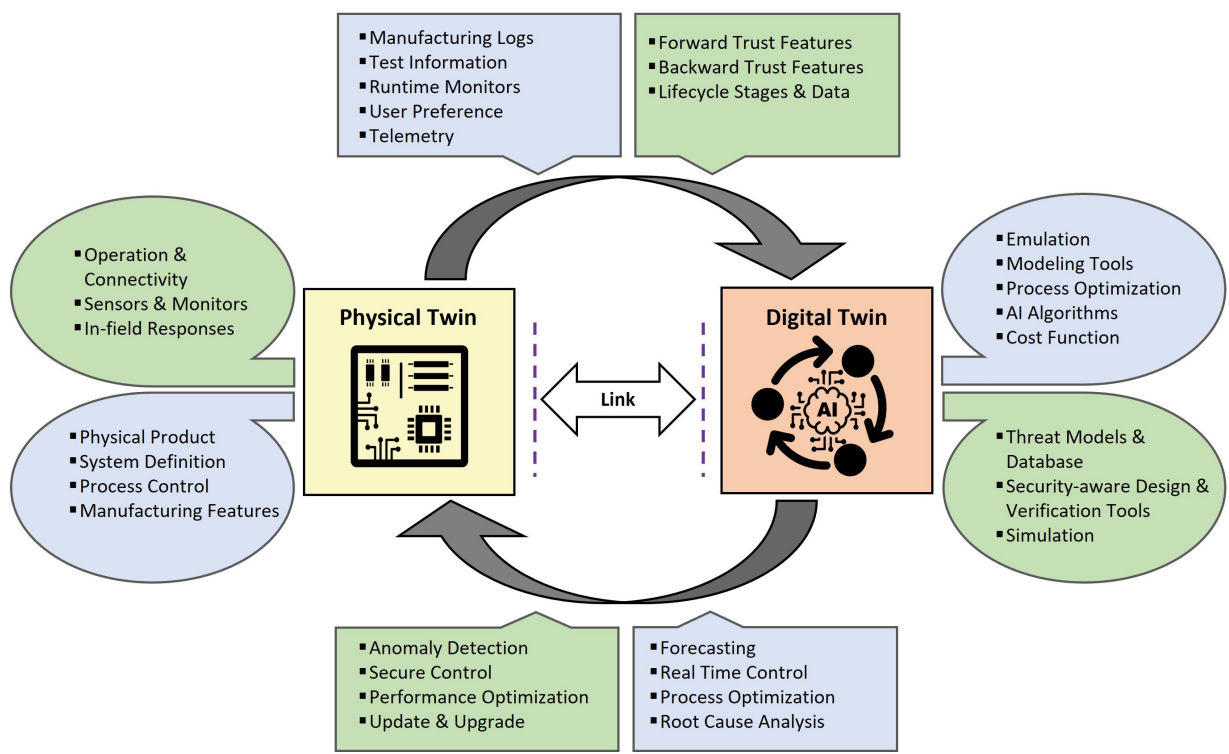}
\caption{Digital twin components, functionalities, and  bidirectional transactions for secure semiconductor lifecycle management. Blue colored boxes and circles indicate components, controls, and transactions of a traditional DT framework focusing product monitoring and process optimization. The green colored boxes and circles indicate those that required to be additionally incorporated into the framework for ensuring end-to-end security and assurance in semiconductor lifecycle. }
\label{figure: highlevel}
\end{figure*}

\section{Introduction}

Amidst rising threats in the supply chain and the ever expanding attack surface, ensuring the security of semiconductor devices across their entire lifecycle has become a challenging and complex endeavor.
Although established practice puts security at the forefront of each stage of the software development lifecycle \cite{khattri2012hsdl}, such efforts to secure the hardware lifecycle is in its infancy due to the unique challenges associated with it \cite{dessouky2019hardfails,sami2021end}. 
The traditional hardware verification and testing methodologies, that focus on functional verification as their primary objective, are often ineffective in detecting security vulnerabilities, which may be introduced through malicious 3PIPs, or security unaware design practices\cite{ray2017system}.
If security vulnerabilities evade detection and verification efforts, they can later be exploited by malicious entities in the supply chain\cite{xu2019electronics,zhang2018chip}.
Unlike software, however, hardware cannot be easily patched, which makes identifying the root origin of the vulnerability in the silicon lifecycle paramount to ensure generational improvement in security assurance. 

Digital twins (DT) have experienced exponential growth in academia as frameworks to monitor, maintain, and control quality and reliability of different products across their different stages of the lifecycle\cite{jones2020characterising}. 
Although originally conceived in \cite{glaessgen2012digital} as a high fidelity digital representation of aerospace vehicles, the concept and definition of the digital twin has evolved to encompass any virtual representation of a physical object, process, or operation which is continually updated by data that is collected across the lifecycle using which it provides optimization feedback on the functionality and control of the physical counterpart (as shown in Figure \ref{figure: highlevel}\cite{fuller2020digital,bao2019modelling,madni2019leveraging}. 
Although DTs have received much recognition as tools of managing product maintenance, fault diagnosis, and monitoring in the aerospace and manufacturing industry\cite{bachelor2019model,martinez2018automatic,kritzler2019digital,qi2018digital}, DTs that consider the full lifecycle are very rare. In fact, Liu et al. report that only 5\% of their reviewed papers on DTs considered the whole lifecycle\cite{liu2021review}.
Furthermore, addressing cyber and hardware security issues that are intertwined with cyber-physical systems utilizing DTs have also been rarely considered\cite{wurm2016introduction}.

Existing solutions proposed in academia and industry to address hardware security concerns also suffer from major limitations when applied in the context of end-to-end secure lifecycle management. 
Firstly, almost all proposed detection and prevention mechanisms are concentrated on very specific threats at specific parts of the lifecycle \cite{guin2016fortis,dabrowski2014towards} with little to no scalability when considering the entire lifecycle and other threat models.

Secondly, although there are multiple detection and prevention methods against hardware attacks including hardware trojans\cite{tehranipoor2014integrated,bhunia2013protection,bhunia2018hardware,xiao2016hardware}, counterfeits \cite{guin2014low,guin2014counterfeit,wang2012representative,guin2014comprehensive,alkabani2007active,huang2015recycled}, information leakage\cite{rajendran2016formal,guo2019qif}, fault injection\cite{wang2021sofi,nahiyan2016avfsm,nahiyan2018security}, and side channel attacks\cite{park2019leveraging,shan2019machine,nahiyan2020script}, the analysis of these methods start with the \textit{a priori} assumption that the defender knows what attack vector is principally responsible for an anomalous behavior.
It is far more likely that the designer or defender would only get to \textit{observe} the anomalous behavior, rather than \textit{knowing} what precise attack vector is causing said behavior. 
For example, from a hardware security perspective, a chip in a system may experience accelerated aging due to many possible reasons. 
It can fail before its intended lifespan because i) it is a recycled chip that was unknowingly used by the system designer, ii) it is a defective chip that was shipped without authorization by an untrusted foundry or a rogue employee working in a trusted foundry\cite{tehranipoor2015counterfeit}, or iii) it experienced accelerated aging due to being taken over by a parametric hardware trojan\cite{shiyanovskii2010process}.

Thirdly, once an attack or vulnerability has been detected, thus far none of the proposed solutions have the capability to trace the lifecycle stage where it originated from.
For example, if an information leakage is detected through formal verification, existing methods cannot infer whether the problem was introduced during high level architecture specification (also known as electronic system level specification) or during the formulation of the logic design (through hardware description languages) of the circuit.
The ability to track down the origin is absolutely vital if we want to facilitate generational improvement in security assurance of the design.

Lastly, there have been suggestions in literature and the semiconductor industry to embed different types of sensors on the chip so that it has a defense against certain attack vectors\cite{kashyap2021silicon,khalid2020simcom}. 
In addition to more area, power, and performance overhead, these approaches are not scalable in the context of emerging future threats. 
New threats and attack vectors are always being developed by researchers and malicious actors. 
As an example, in the initial years of hardware trojan research, it was frequently assumed that hardware trojans need to be activated by rare signals and node to avoid detection. 
However, researchers have since demonstrated that it is possible to design hardware trojans that do not need be triggered by rare events yet easily escape traditional testing and verification efforts\cite{lin2009moles}.
It is not feasible to keep continually adding new sensors to a design to tackle new threats as they emerge.

Therefore, we argue that without a comprehensive framework, such as the one we are proposing in this paper, security assurance in the semiconductor lifecycle would only be limited to partial effectiveness with severely limited scalability (no matter how robust individual detection algorithms or protection mechanisms are).
A DT with bidirectional data flow and feedback, as shown in Figure \ref{figure: highlevel} between the real world and the virtual presents a suitable concept around which data collection and analysis tools, algorithms can be leveraged to build a comprehensive framework to provide security assurance across the whole lifecycle by addressing each of the aforementioned challenges. 
The main contributions of our work lie in the following:
\begin{itemize}
    \item We propose a digital twin framework that can provide security assurance across the entire lifecycle by considering the potentially malicious supply chain entities and vulnerable cycle phases. DT deconstructs the problem by analyzing causal relationships between available data and hardware security vulnerabilities. Thus, instead of addressing one or two attack vectors, DT provides a scalable methodology to combat potentially all possible hardware attack vectors. 
    \item Our proposed methodology theorizes the use of data that is already being gathered by the traditional process flows in the silicon lifecycle. Consequently, adoption of our framework incurs no hardware overhead and offers a promising prospect of being seamlessly integrated into existing flows.
    
    \item We define the feedback from the DT to the physical world in terms of two functionalities: namely, \textit{Backward} and \textit{Forward Trust Analyses}, respectively. \textit{Backward Trust Analysis} provides traceability through root cause analysis of observed anomalous behavior in device security policies at any stage of the lifecycle. 
    We demonstrate how artificial intelligence (AI) algorithms can be used to perform reliable root cause analysis in the hardware security domain. To perform this root cause analysis, we explore three different statistical relational learning algorithms, namely Bayesian Networks, Hidden Markov Models and Markov Logic Networks, by each of which causal inference can be performed. Additionally, we demonstrate how they can be adapted to the problems of silicon lifecycle security.

    \item The dichotomy of security assurance is that on one hand, as time passes, novel threats emerge that circumvent existing protection and detection measures. On the other hand, the collective understanding of these newly emerging threats calcify, which gives rise to better performing prevention and detection methodologies. Through our proposed framework, we demonstrate how it can be made scalable and continually updatable, which in turn can preserve applicability of its ability for root cause analysis (even against unforeseen threats). This scalability and adaptability is what we refer to as \textit{Forward Trust Analysis}. 
\end{itemize}

The rest of this paper is organized as follows: Section \ref{sec: prelim} provides a literature review of data driven approaches in silicon lifecycle management and various DT applications proposed in cybersecurity and lifecycle management. Section \ref{sec: motivation} provides two running motivating examples which are used throughout the paper to illustrate usability of the proposed DT. Section \ref{sec: taxonomy} presents the entire silicon lifecycle with an emphasis on available data throughout different stages of the lifecycle. Section \ref{sec: HAV} provides a basic introduction to three relevant hardware attack vectors, challenges of providing security assurance against them, and also provides insight into how data from different lifecycle stages are related to security vulnerabilities associated with the scenarios. Section \ref{sec: DT structure} elaborates on the structure of the proposed DT framework. Existing challenges and future research directions in implementing the proposed DT are laid out in section \ref{sec: challenges}. Finally, Section \ref{sec: conc} provides a summary of the discussions and concluding remarks.
\section{Preliminaries}
\label{sec: prelim}
\subsection{Digital Twin at a Glance}
\label{subsec: DT}

The concept of DT has evolved to encompass many different definitions\cite{fuller2020digital}. 
Some authors have put strong emphasis on the simulation aspect of DTs, whereas others have argued for clear definition of three aspects (physical, virtual, and connection parts) as the criterion for a framework to be called a digital twin\cite{tao2018digital}. 
We use the definition provided by Madni\cite{madni2019leveraging} in context of the lifecycle management of products to illustrate the different components of a DT system in Figure \ref{figure: highlevel}.
At the core of a DT is the collection of sensor, simulation, emulation, and preliminary analytics data that are gathered across a physical device's lifecycle. 
The physical process, or device, is also referred to as \emph{Physical} twin. 
The twins are housed within environments that are referred to as physical and virtual environments, respectively.
The \emph{Digital} counterpart is formed by continually updating the database hosted in the virtual environment.
The DT is capable of providing intelligent feedback (e.g., forecasting, optimization of parameters, root cause analysis, real time control) to the physical world through a combination of simulation, emulation, data analytics, and AI modeling. 
The communication links between the physical and virtual environments are also essential components of the DT.
It is imperative to note here that a digital twin is not merely a single algorithm or a single technology\cite{liu2021review}, but rather a framework around which a systematic methodology can be built to combat product lifecycle issues.
For security assurance across the whole lifecycle of a semiconductor device, it should be noted that having only the traditional components and transactions are not sufficient since they do not necessarily offer security-aware features. 
Hence, additional transactions and functionalities are required as indicated in green boxes and circles in Figure \ref{figure: highlevel}.
It also calls for advanced machine learning, statistical relational learning, and other data analytic-related algorithms to gleam insight from gathered data.
The methods, algorithms, structure and contingencies required to realize these additional components and transactions are discussed throughout the rest of the paper.

\subsection{Digital Twin for Lifecycle Management and Cybersecurity}
\label{subsubsec: DT for LM}
\begin{table*}[!h]
\caption{Digital Twin Applications Suggested in Literature for Lifecycle Management and Cybersecurity}

\label{table:related works}
\centering
\begin{tabular}{p{0.18\textwidth} p{0.24\textwidth} p{0.50\textwidth}}
\hline
\textbf{Paper} & \textbf{Application Area} & \textbf{Comments} \\
\hline
Bitton et al.\cite{bitton2018deriving} & Cybersecurity of Industrial Control & Proposed the use of DT to overcome the limitations of existing network \\ 
  & Systems (ICS) & penetration testing when applied to industrial SCADA systems \\
\hline
Lou et al.\cite{lou2019idea} & Cybersecurity of ICS & Demonstrated the use of DT to address security issues of a refueling machine \\
\hline
Balta et al.\cite{balta2019digital}& Process management & Proposed a DT for anomaly detection and process monitoring of the fused deposition modelling AM process.\\ 

\hline
Eckhart et al.\cite{eckhart2018DT}& Network and CPS security & Proposed a CPS twinning system where states of the physical systems are mirrored through the DT that can incorporate security enhancing features, such as intrusion detection.\\ 
\hline
Saad et al. \cite{saad2020implementation}& Network and grid security & Illustrated a DT's capability in providing security against false data injection, Distributed-Denial-of-Service (DDoS) and network delay attacks in microgrids.\\
\hline
Li et al. \cite{li2017dynamic}& Product lifecycle management & Proposed fault diagnosis and prognosis technique in aircraft wings through a dynamic Bayesian Network driven DT.\\
\hline
Sleuters et al. \cite{sleuters2019digital}& System management & Proposed a DT to capture the operational behavior of a distributed IoT system.\\
\hline
Wang et al. \cite{wang2021digital}& Smart manufacturing & Discussed how a DT may be used for intelligent semiconductor manufacturing.\\
\hline
Jain et al. \cite{jain2019digital}& System management & Proposed a DT to offer real time analysis and control of a photovoltaic system.\\
\hline
Xu et al. \cite{xu2019digital}& Process management & Demonstrated a DT that offers real time diagnosis and predictive maintenance of a car-body side production line.\\
\hline
Kaewunruen et al. \cite{kaewunruen2019digital}& Operational lifecycle management & Proposed a DT for sustainable management of railway turnout systems.\\
\hline
Heterogeneous Integration Roadmap 2021 Ed. \cite{HIR2021}& Reliability management for Semiconductor & Briefly discussed possible DT prospects for reliability management of semiconductor devices.\\
\hline
Alves et al.\cite{alves2019digital}& System management & Developed a DT to monitor and control water management in agricultural farms\\
\hline
Tchana et al.\cite{alves2019digital}& Operational lifecycle management & Developed a DT to address operational issues in linear construction sector. \\
\hline
\end{tabular}
\end{table*}
Fault diagnosis or root cause analysis as a core functionality of digital twins in context of product lifecycle management and industrial production has been explored in several works\cite{xu2019digital,jain2019digital,sleuters2019digital}.
DTs have been demonstrated to be applicable for lifecycle management in agricultural\cite{alves2019digital} and Building Information Modelling (BIM) systems\cite{tchana2019designing}.
The existing literature on DT for cybersecurity focuses mainly on network and software security\cite{pokhrel2020digital}. 
The focus has been on identifying intrusion\cite{eckhart2018DT} or false data injection attacks in an industrial setting. 
Bitton et al. proposed the use of a DT specified \emph{automatically} from a rule set derived from tests and a so called problem builder derived the constraints by solving a non-linear maximization problem\cite{bitton2018deriving}.
In a similar setting of an ICS, DT has been used to resolve security issues associated with a refueling machine\cite{lou2019idea}.
Saad et al. addressed attacks from potentially multiple coordinated sources on a networked micro grid\cite{saad2020implementation}.
The reader should note that these approaches only consider specific type of control systems, not the security issues associated with the entire lifecycle.
Lifecycle management of products, especially security management, requires additional capabilities, considerations and bidirectional transactions.

A high-level formulation of digital twins for semiconductor reliability can be found in \cite{HIR2021}.
Reliability concerns are inherently limited to considering a subset of the lifecycle as vast majority of semiconductor reliability concerns originate from fabrication and packaging processes. 
Another discussion of digital twins in context of the semiconductor fabrication process can be found in\cite{wang2021digital}.
Again, this discussion is limited to only one phase of the lifecycle in context of smart manufacturing and not related to security concerns.
The current dominant trend in academia, which is evident in this brief literature review section as well, is to utilize digital twin for systems which are almost exclusively manufacturing systems or processes.
We buck that trend in our paper by showcasing how digital twins can contribute significantly in secure lifecycle management as well.

An overview of the papers discussed in the preceding can be found in Table \ref{table:related works}.

\subsection{Hardware Security and Trust}
\label{subsec: HWST}

Over past several decades, hardware of a computer system has traditionally been considered as the emph{root of trust} to guard against attacks on the software running on the system. 
The underlying assumption here is that since hardware is less easily malleable than software, it is likely to be robust and secure against different types of attacks \cite{bhunia2018hardware}. 
However, emerging hardware attacks that exploit intralayer and crosslayer vulnerabilities have propelled hardware security as a widely researched topic.
The recent proliferation of reported attacks on hardware is not surprising given how the business model of the semiconductor industry has evolved over the course of past few decades. 
Previously, all stages associated with bringing a semiconductor chip to the market (namely design, fabrication, test, and debug) were handled by a single entity. 
To address aggressive time-to-market demands and profitability concerns, the global semiconductor industry has gradually adopted a horizontal model of business wherein each previously mentioned stage may be handled by completely different entities often situated in different parts of the globe. 

Several of the IPs used in a typical SoC are procured from third-parties across the globe \cite{rostami2014primer}. 

Today, most SoC design houses are ``fabless," meaning that they do not own a foundry to physically fabricate the chips that they design. 
They rely on a foundry and a third party to respectively fabricate and distribute the chips for them.
The consequence of this distributed manufacturing and supply process is that the security of the designed SoC may become compromised through malicious modification by multiple entities across the entire supply chain.
This raises the issue of trust between multiple entities referred to as hardware trust. 
    \begin{figure*}[!h]
    \centering
    \includegraphics[width=7in]{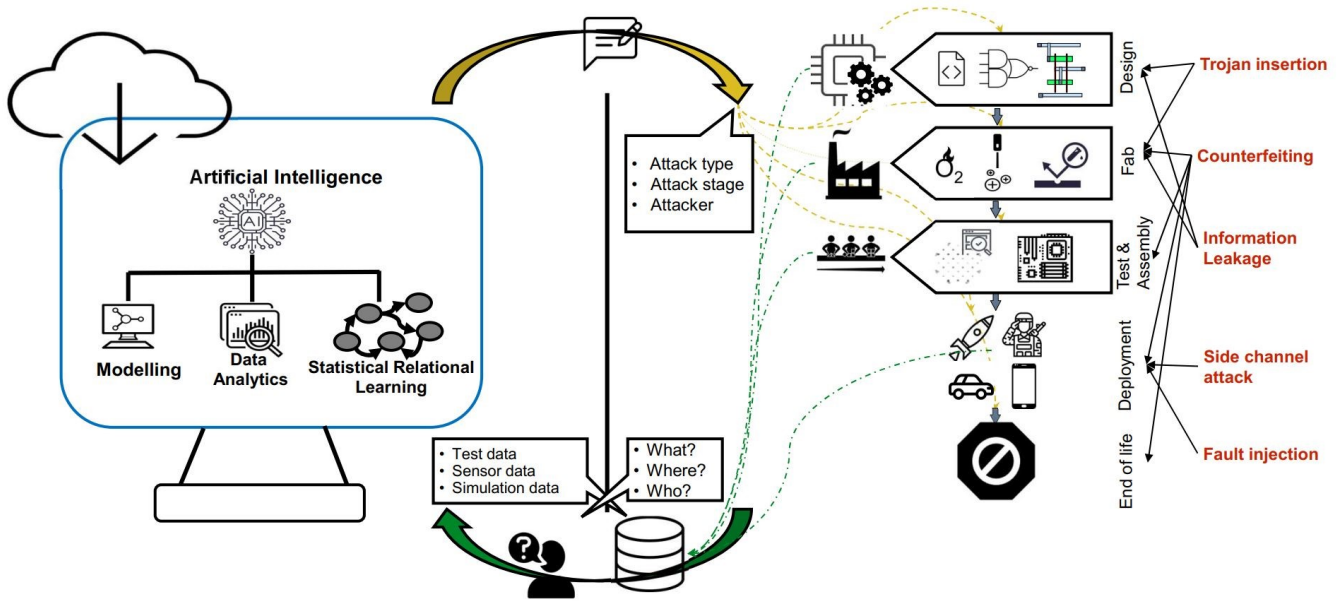}

    \caption{A high level overview of an AI driven DT framework adapted for security assurance in semiconductor lifecycle against various hardware attacks (highlighted in red). The virtual environment is the cloud containing a digital representation of the design and lifecycle information of the physical semiconductor device(s). The DT would be realized with a combination of computer modeling, data analytics and AI algorithms. To address challenges of hardware security and trust, the DT would need to analyze the test, simulation/emulation, and sensor data uploaded to it throughout the lifecycle. With properly defined threat models, the DT would not only infer the underlying attack vector, but also the lifecycle stage where it occurred. An iterative loop of data, control, and reasoning will allow existing threat detection with precision (enabling backward trust) as well as provide insights for preventing emerging threats and zero-day attacks (enabling the forward trust).}
    \label{figure: detailedhigh}
    \end{figure*}
Hardware attacks can take place in the form of a malicious modification of circuit (hardware trojan), stealing of the IP by the foundry, recycling and remarking of chips, physical tampering\cite{asadizanjani2021physical,rahman2018physical}, reverse engineering\cite{quadir2016survey,botero2021hardware}, and side channel attacks by end users. 
These attacks might be carried out by different actors in the supply chain who may have different goals. 
In addition to these attacks, various vulnerabilities might be introduced unintentionally in the design, such as the leakage of a security critical asset through an output port or to an unauthorized IP. 
The possible hardware attacks and the stages in which they might occur are highlighted in red on the right side of Figure \ref{figure: detailedhigh}.
These attack vectors have highly varied associated threat models, characteristic symptoms, and detection methodologies. 

\subsection{Data-driven Approaches for Assuring Quality, Reliability, and Security in Semiconductor}

The basic building blocks of a digital twin (i.e., data collection and analytics) are already an indispensable part of existing flows in traditional semiconductor lifecycle; however, data driven approaches that leverage this sizeable amount of data to manage the whole lifecycle have rarely been reported.

Data obtained from these steps can be analyzed to provide assurance to broadly three aspects of the semiconductor lifecyle, namely quality, reliability, and security. 
As such, a digital twin framework can be constructed to enhance each of these aspects without drastic modification of its existing design and process flows.

It should be noted that there is a fundamental difference between product lifecycle management and security assurance through lifecycle management.
While the former is concerned with satisfying the functional requirements of a product and diagnosis of the underlying causes upon failure to do so, the latter is concerned with preserving the desired security properties of a system against attacks or unintentional mistakes of the designer.

In the last two decades, malicious modifications, vulnerabilities, and attacks on hardware have been extensively reported in literature and the press \cite{AMD,TPM,fournaris2017exploiting, king2008designing,lee2020off}.

To the best of our knowledge, there have only been two approaches in the silicon lifecycle management in literature that attempt to provide security assurance to the lifecycle.
In \cite{kashyap2021silicon}, authors present their Synopsys SLM platform to assure quality, reliability, and security across the lifecycle.
The proposed platform uses proprietary data engines to gain \emph{actionable insights} to address various design and manufacturing issues. 
Although authors claim that the analytics engines can be used for bolstering security defenses, there is no clear guideline provided on what data items are related to security vulnerabilities and how these relationships can be leveraged to defend against different types of security threats.
Inspired by similar practices in software domain, the authors present a hardware secure development lifecycle (HSDL)\cite{grand2004practical} composed of five phases to identify and mitigate security issues as early as possible in the lifecycle.
However, the proposed approach is a general pointer on \textit{what} steps to follow for secure hardware development without specifics on \textit{how} to achieve them through a singular framework. 
In \cite{dabrowski2014towards}, traceability for hardware trojans is provided through a unified framework, however, it only does so for a specific hardware threat vector.

\section{Digital Twin for Securing Semiconductor Lifecycle Management: Problem Definition and Motivating Examples}
\label{sec: motivation}

In the hardware security domain, academia has proposed many different algorithms and testing methodologies to detect different types of hardware attacks.
Also, many proposals called design for security (DfS) approaches have been inspired by established design for testing (DfT) practices, which advocate for embedding different sensors into a chip or leveraging data from existing chips to better prevent attacks. However, the challenge is that the device is more likely to exhibit an anomalous behavior during its operation or when subjected to a test, thus it is up to the defender to understand why this behavior is occurring. 
As semiconductor industry has gradually shifted from a vertically integrated business model to a globally distributed one, there can be multiple possible explanations for a single anomalous behavior as there are many untrusted entities in the supply chain. 
Vast majority of existing literature on defense against hardware attacks have the underlying assumption that the attack vector is already known and detection or prevention methods against that attack vector need to be developed. 
This assumption makes sense if the threat model under consideration makes appropriate assumptions. 
In context of the whole lifecycle though, such restrictive threat models do not apply. 
A naive solution might be to put preventive measures in place on the chip to address all possible attacks; however, as the sensors and circuitry required are different, the performance penalty and hardware overhead for doing so would be unacceptable.

As a motivating example, let us consider three different scenarios.
In the first two cases, the semiconductor device is a chip designed by a fabless design house.

\paragraph{\textbf{Scenario 1}}
\label{scene1}

The design house receives customer feedback that a certain number of chips designed by them is experiencing accelerated aging.
For the sake of focused discussion on security assurance, let us also assume that the accelerated failure is not a reliability issue as the design passed through all reliability checks during the design phase.
In this scenario, the designers and the CAD tools used by the design house are considered trusted.
Now, the design house has to consider at least three possible explanations behind this behavior:
\begin{itemize}
    \item [i.] The failing chips are recycled or remarked chips that got resold as after they had reached their end of life.
    \item [ii.] The failing chips are out-of-spec or defective chips that should have failed the burn-in or wafer probe test during test and assembly. A rogue employee in the foundry or potentially the untrusted foundry itself is shipping some of the chips that failed these tests.
    \item[iii.] The failing chips are infested with a process reliability based trojan inserted by a rogue employee or the untrusted foundry.

\end{itemize}

\paragraph{\textbf{Scenario 2}}
\label{scene2}
Infield testing such as JTAG testing and Built-in-Self-Test (BIST) are often carried out after deployment to debug performance anomalies.
In this scenario, let us assume that during such testing it is found that a confidential asset such as a secret key can be observed through the debug ports.
Similar to scenario 1, there might be multiple possible explanations each of which arise from either security vulnerabilities introduced or attacks performed earlier in the lifecycle. 
We assume that the CAD tools can extensively verify information leakage flows.
The possible explanations for this behavior are as follows:
\begin{itemize}
    \item [i.] Designer overlooked the proper implementation of security policies while writing the hardware description code or even earlier at high level design specification stage.
    \item [ii.] A malicious information leaking hardware trojan was introduced in the circuit in the design phase through 3PIPs or inserted by the untrusted foundry.
\end{itemize}

Given these two scenarios, backward trust functionality of our proposed DT functionality will assign a probable provenance to the observed anomalous behavior through root cause analysis.
Backward trust also entails identifying the possible causes of an observed anomalous behavior in the first place.
This functionality is illustrated in Figure \ref{figure: detailedhigh} where the queries driving backward trust analysis are highlighted: what type of attack it is, where in the lifecycle it originated from and who was responsible. 
The DT framework also will facilitate forward trust by ensuring that it is adaptable to future threats insofar as their successful identification and application of root cause analysis are concerned.

\section{Semiconductor Lifecycle Data for Digital Twin Mapping}
\label{sec: taxonomy}

\begin{sidewaysfigure*}[htbp]
\centering
\subfloat{\includegraphics[width=9.25in]{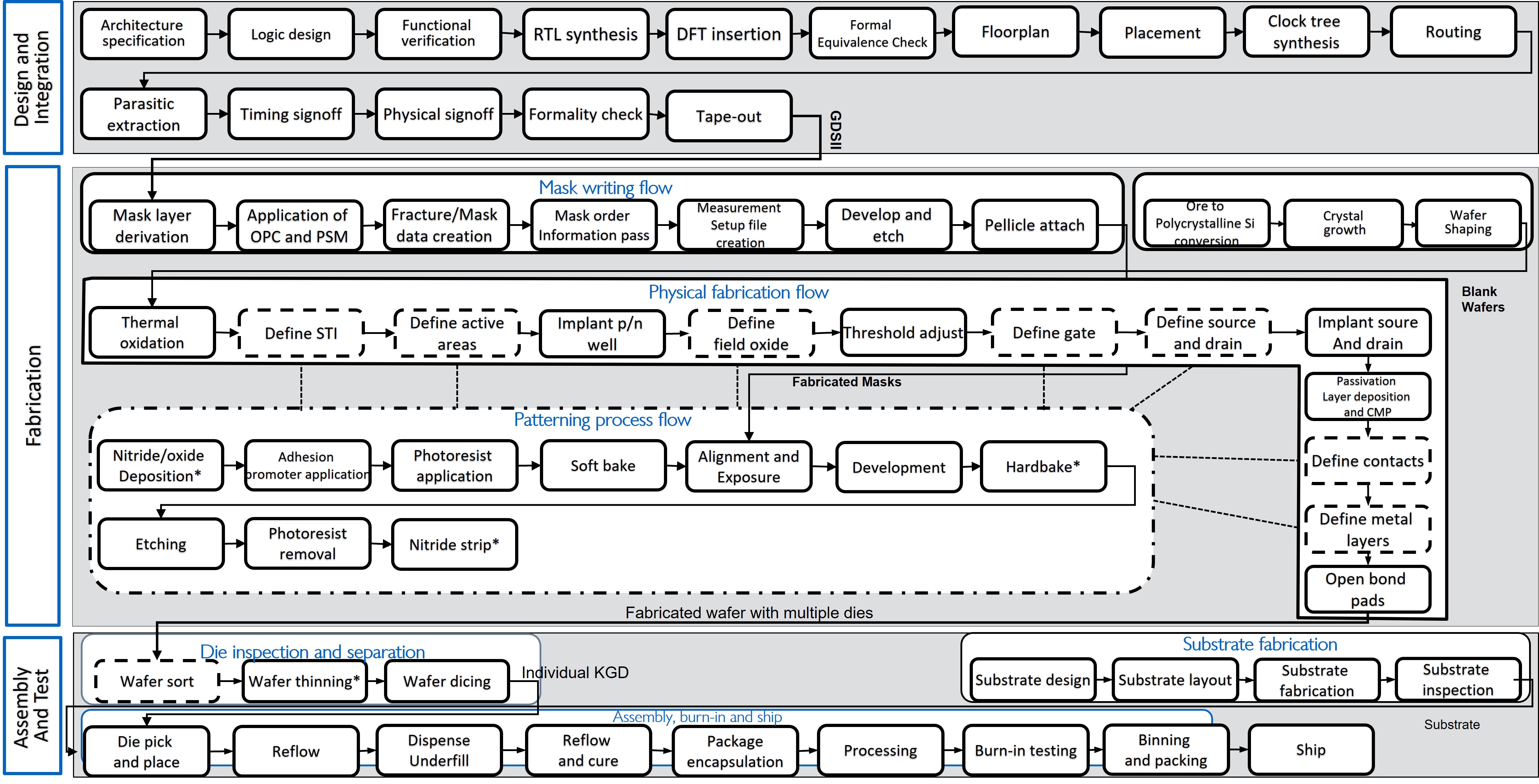}}%
\caption{A high-level overview of the first three stages of semiconductor lifecycle: Design, Fabrication, and Assembly \& Test. The sub-stages represented with dashed line boxes indicate that they themselves can be dissolved into further sub-stages. In the fabrication stage, broadly the same processes are repeated to pattern different areas of the fabricated chip. This repetition of processes is denoted by a dashed line connecting the \emph{patterning process flow} box with various sub-stages. The exact fabrication processes may vary depending upon the technology used by the foundry. The assembly and packaging process shown here is for Ball Grid Array (BGA) packages. For a different packaging technology (e.g., 3D and Quad Flat Packaging) the processes may be slightly different.}
\label{figure: substage}
\end{sidewaysfigure*}

At the heart of every digital twin, there is bidirectional data flow between the physical and virtual environments. 
Data can be exchanged either as it is collected or after preliminary analytics has been performed on it.
As a semiconductor device moves through various phases of its lifespan, the tools and software that are used to design it, the machinery that are used to manufacture it as well as the tests that are carried out to ensure its proper operation generate a huge volume of data. 
The reader can consult Figure \ref{figure: substage} to get a glance of the numerous steps a chip has to go through before it is ready for use for in-field applications.
The life of a semiconductor chip  ends at the recycle facility. 
In between, the chip is fabricated by a foundry, assembled and tested and distributed to the market. 
Additional design features have to be added to each designed chip to reduce effort and complexity of testing and debugging. 
These design for test (DfT) and design for debug (DfD) infrastructures are sometimes outsourced to third parties.
Each of these stages of the lifecycle consists of several sub-phases; a high level overview showcasing every sub-phase for the design, fabrication, test and assembly stages in Figure \ref{figure: substage}.

A discussion on DT for secure semiconductor lifecycle is impossible without an understanding of the lifecycle stages and available data therein. 
Therefore, this section presents a brief description of each of the lifecycle stages along with an emphasis on the gathered data.
At each stage of the lifecycle, industrial practice dictates the extensive collection and analysis of data to ensure reliability as well as satisfactory performance.
Security is an afterthought in most cases although the rising threats in the global supply chain necessitate that the collected data be used and analyzed for security assurance as well.
Academia has suggested various secondary analysis on the available data that may be used for that purpose. 
For a detailed reference to how the collected data may be used to perform security assurance evaluations, the reader is advised to consult section \ref{subsection: security}. 
\subsection{Pre-silicon Design Stage}
In the design phase of a semiconductor chip, a blueprint of the chip to be fabricated is prepared and delivered to the foundry in the form of a GDSII file\cite{rahman2021security}. 
The design must satisfy all specifications and perform desirably under all operating conditions and constraints (in terms of power, area, timing etc.) of interest. 
The design phase itself can be further subdivided into multiple sub-phases, all of which form a sequential flow starting from architecture specification and ending at tape-out\cite{ahmed2021quantifiable}. 
Today, most of the following sub-phases in the process are automated using a combination of commercially available and open source software.
\begin{itemize}
    \item [a.] \textbf{Architecture specification:} At this phase, a high level description of the circuit to be designed is prepared taking various trade-offs and customer feedback into consideration. Vast majority of circuits ship with a specification sheet that has detailed data on intended functionality, user guide on how to use the setup and use the hardware debugging features, data on important electrical, physical and architectural properties such as operating temperature, clock speed, memory size, interface protocols etc.
    
\begin{table*}[!h]
\caption{Data Available at Design Stage}
\label{design_table}
\centering
\begin{tabular}{p{0.18\textwidth} p{0.18\textwidth} p{0.28\textwidth} p{0.28\textwidth}}
\hline
\textbf{Design sub-stage} & \textbf{Output Data files} & \textbf{Primary information/data} & \textbf{Secondary analysis data}\\
\hline
Arch. Specification & Specification datasheet & Description of the intended functionality, operating limits, specific protocols and technologies used & - \\
\hline
Logic Design &  RTL & Behavioral and/or structural description of the circuit & Branching probability, relative branching probability \cite{choo2020register}, side channel leakage assessment score \cite{he2019rtl}\\  
\hline
Functional verification &  SAIF, VCD & File metadata, timescale, variable type and bit-length, identifier codes & Coverage metrics \cite{fallah2001occom,serrestou2007functional},  \\
\hline
RTL to gate level synthesis &  Netlist,  SDF & List of nodes and their interconnections, delay information & Controllability and observability of nets \cite{goldstein1980scoap}, rare nets \cite{waksman2013fanci,chakraborty2009mero} \\
\hline
DfT insertion &  DfT inserted netlist, SDF & Same as RTL to gate level synthesis & Same as RTL to gate level synthesis \\
\hline
Formal equivalence and model checking &  Equivalency report & No. of primary outputs, and points in the circuit that are equivalent & Test pattern generation for hardware trojan detection \cite{cruz2018hardware}, secure information flow verified 3PIP cores \cite{rajendran2016formal}, Code, functional and toggle coverage \cite{nahiyan2017code} \\
\hline
Floorplan and placement & Floorplan and placement db & Size and coordinates of the IO pads, IP blocks and power network & Chip temperature \cite{xu2017fast}, power supply noise \cite{cong2004area}, overall electromagnetic radiation \cite{ma2020security} \\
\hline
Clock tree synthesis &  Clock tree db & Clock skew and latency & Unused spaces in the layout \cite{xiao2013bisa} \\
\hline
Routing &  Routed db (e.g., DEF ), post place and route netlist & Detailed location and geometry of interconnects & Wire length adjustments required for suppressing electromagnetic leakage \cite{ma2020security} \\
\hline
Parasitic extraction &  SPEF & Parasitic resistance and capacitance of nets and interconnects & - \\
\hline
Power, timing, and physical signoff &  Timing, power and physical signoff reports & Static, dynamic and leakage power, delay information, setup and hold time, and no. of DRC and LVS violations. &  Path delay fingerprinting for hardware trojan detection\cite{salmani2009new}\\
\hline
Tape out & GDSII & Binary representation of layout geometries & - \\
\hline
\end{tabular}
\end{table*}
    \item [b.] \textbf{Logic design:} Hardware Description Languages (HDLs) such as Verilog, VHDL, SystemVerilog etc. are used to describe and capture the specifications determined in the previous step. 
    The code written at this stage also known as Register Transfer Level (RTL) code.
    The code itself may describe the behavior and/or the structure of the circuit in a specialized language.
    \item[c.] \textbf{Functional verification:} Next, the written RTL code is tested against the specification to verify whether it has successfully captured the intended behavior and functionality.
    The results of functional verification, also known as logic simulation, are often stored in a Value Change Dump (VCD) file which contains information on the sequence of value changes in different signal variables with respect to time along with the file meta-data, definition of signal variables and timescale \cite{vcd}. 
    \item[d.] \textbf{RTL to gate level synthesis:} The RTL code is then synthesized to produce a schematic of the circuit in terms of constituent logical gates. This process is entirely automated with help of commercially available tools. The synthesized netlist is a description of the nodes in the circuit along with the interconnection between these nodes. 
    \item[e.] \textbf{DfT insertion:} As mentioned previously, the complexity of modern VLSI circuits necessitates the inclusion of additional features in the design for increased testability of designs. DfT insertion step has similar outputs to RTL to gate level synthesis step.
    \item[f.] \textbf{Formal equivalence check:} The design is verified by formal assertions in the form of logical and mathematical properties at this step. It provides a mathematical proof of functional equivalence between the intended design and the synthesized netlist. At the end of the verification, the designer is informed of the no. of points in the design that are equivalent to the intended functionality of the design.

    \item[g.] \textbf{Floorplanning \& placement:} Floorplanning refers to the organization of circuit blocks within small rectangular spaces of the available space. The precise location of the I/O pins, power and clock distributions are determined in the placement step. 
    \item[h.]
    \textbf{Clock tree synthesis:} Clock tree synthesis step ensures the even distribution of the clock to all sequential elements in the design, satisfaction of all timing constraints as well as minimization of skew and latency by clock tree building and balancing. 
    \item[i.] \textbf{Routing:} In the routing step, the myriad of interconnects that connect different cells with each other as well as the individual gates within each cell get outlined.  
    \item[j.]\textbf{Power, timing and physical signoff:} Physical signoff involves the verification of the physical design performed in the last four steps against technology node defined design rules. Timing and power signoff verifies the physical design against timing and power requirements. At the end of verification, the designer has detailed information on whether important circuit parameters such as hold and setup time, dynamic, static and leakage power, and interconnect and pat delay meet design requirements. 
    \item[k.]\textbf{Tape out:} The verified design is then shipped out to the foundry in the form of a GDSII file. 
\end{itemize}

As mentioned previously, automation effort in the design process is achieved by the use of commercially available software. 
These software collect and analyze data at each step to optimize performance and reliability of the design. 
The data available from each stage is often stored in the form of different software file formats. 
In literature, various types of secondary analysis have been proposed which can be performed on each of these files to derive secondary data of interest.
For example, the RTL file describing the behavioral specification of the circuit can be analyzed to get information about branching probability, control flow graph (CFG), data flow graph (DFG) which in turn may be used for security and performance optimization purposes \cite{forte2017hardware, choo2020register}. 

A summary of available data from these sub-stages of design phase can be found in Table \ref{design_table}. The readers are advised to note that this table is not exhaustive; particularly, the data file formats and available secondary data may change depending on the software being used and the type of analysis being performed respectively.

\subsection{Fabrication and Manufacturing Stage}

CMOS fabrication is an extremely sophisticated process in which the exact steps followed depends on a large variety of factors including the technology node, the operating conditions that the chip is expected to experience, performance as well as cost considerations, the device or application in which it is to be used on, and many more. 

The most important technique during the fabrication process is photolithography which refers to a process to transfer an etched pattern from a chromium or quartz plate- called a photomask- onto a silicon wafer using light or electromagnetic radiation\cite{mack2008fundamental}.

The middle box in Figure \ref{figure: substage} shows sub-stages of an example fabrication process\cite{May2003fabrication,Campbell2008fabrication}. 
This section describes the CMOS VLSI fabrication process in brief and also provides a summary of data collected from various equipment on the manufacturing floor.

\subsubsection{Mask writing flow}
\label{subsubsec:maskwrite}

    The process of manufacturing photomasks is known as mask writing. 
    The goal of the mask writing process is to transform the GDSII file, which is a binary file format, to a format which is understood by mask writing tools \cite{cobb2002hierarchical} as well as to break the complex shapes present in the GDSII into simpler polygons.
    A data preparation step starts the mask writing process by performing graphical operations such as using Boolean mask operations to derive mask layers that were not part of the original input data. 

    Next, to facilitate printing of features that are much smaller than what would be possible for a particular wavelength of incident light, the geometrical shapes that are present in the GDSII need to be augmented by applying different resolution enhancement techniques (RETs) such as: optical proximity error correction (OPC) for nearby features, phase shifting features, scattering bars etc.

    For verification and metrology, the masks contain barcodes and mask IDs. 
    The RET applied mask data along with these additional data is then `fractured' into simpler polygons that can be handled by the mask writing tools. 
    The consequence of addition of all these data to the original design data is that the fractured file size is often several times more than the original GDSII \cite{BACUS,schulze2002gds}.  
    The mask data then needs to be passed to a mask shop, typically outside the foundry, who will manufacture these masks. 

    Often instructions regarding how and where to carry out these measurements are contained in a specific file, known as measurement setup file, which are loaded into the tools. 
    Once this file has been created, the masks are then physically fabricated using similar processes to those that are used in fabrication of the chip itself. 
    For the sake of brevity, the discussion of these processes are only presented once, later in this section. 
    The manufactured masks are extremely expensive costing up to millions of dollars. 
    As such, they are encased in a protective membrane called pellicle to protect against erosion, dust particle adherence, and other mechanical damage.

\subsubsection{Physical fabrication}
\label{subsub:physical}

\begin{sidewaysfigure*}[htbp]
\centering
\subfloat{\includegraphics[width=7in]{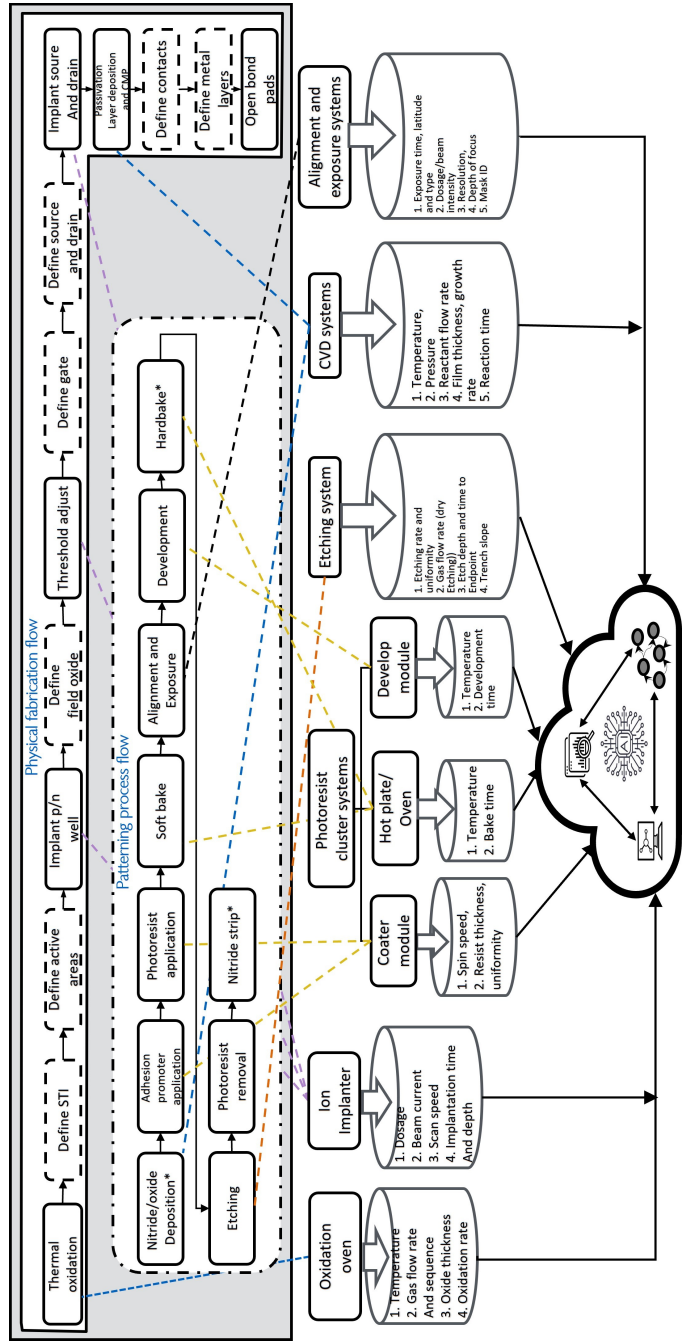}}%
\caption{Representative fabrication steps and data obtained therefrom. The colored dashed lines connecting equipment with steps indicate that the equipment is used for that step. On the other hand, the black dashed lines around a box indicate that particular step contains many sub-steps which in turn are shown in the `patterning process flow" box. Many of the steps are carried out in the same equipment on the foundry floor and therefore, generate the similar type of data. As the data is dependent on the particular tools being used, the exact list of data obtained may be different from what is shown, depending on factors such as equipment vendors, process recipie, etc.}

\label{figure: fab}
\end{sidewaysfigure*}

The polycrystalline Electron Grade Silicon(EGS) is processed through an apparatus called crystal puller to create the silicon ingot which is then mechanically shaped to a closed disc shaped wafer \cite{May2003fabrication}. 
These blank wafer preparation steps are shown in the top right corner of the middle box in Figure \ref{figure: substage}. 
Using the masks prepared in mask writing flow and the blank wafers, the physical fabrication of the circuit commences. 
Again, the reader should be advised that the process flow that is presented in Figure \ref{figure: substage} or \ref{figure: fab} and described in greater detail in this section is representative; not exhaustive. 
For circuits which are fabricated using a different for example, silicon-on-insulator (SOI) technology, the precise steps may differ.
\begin{figure*}[h]
\centering
\subfloat{\includegraphics[width=7in]{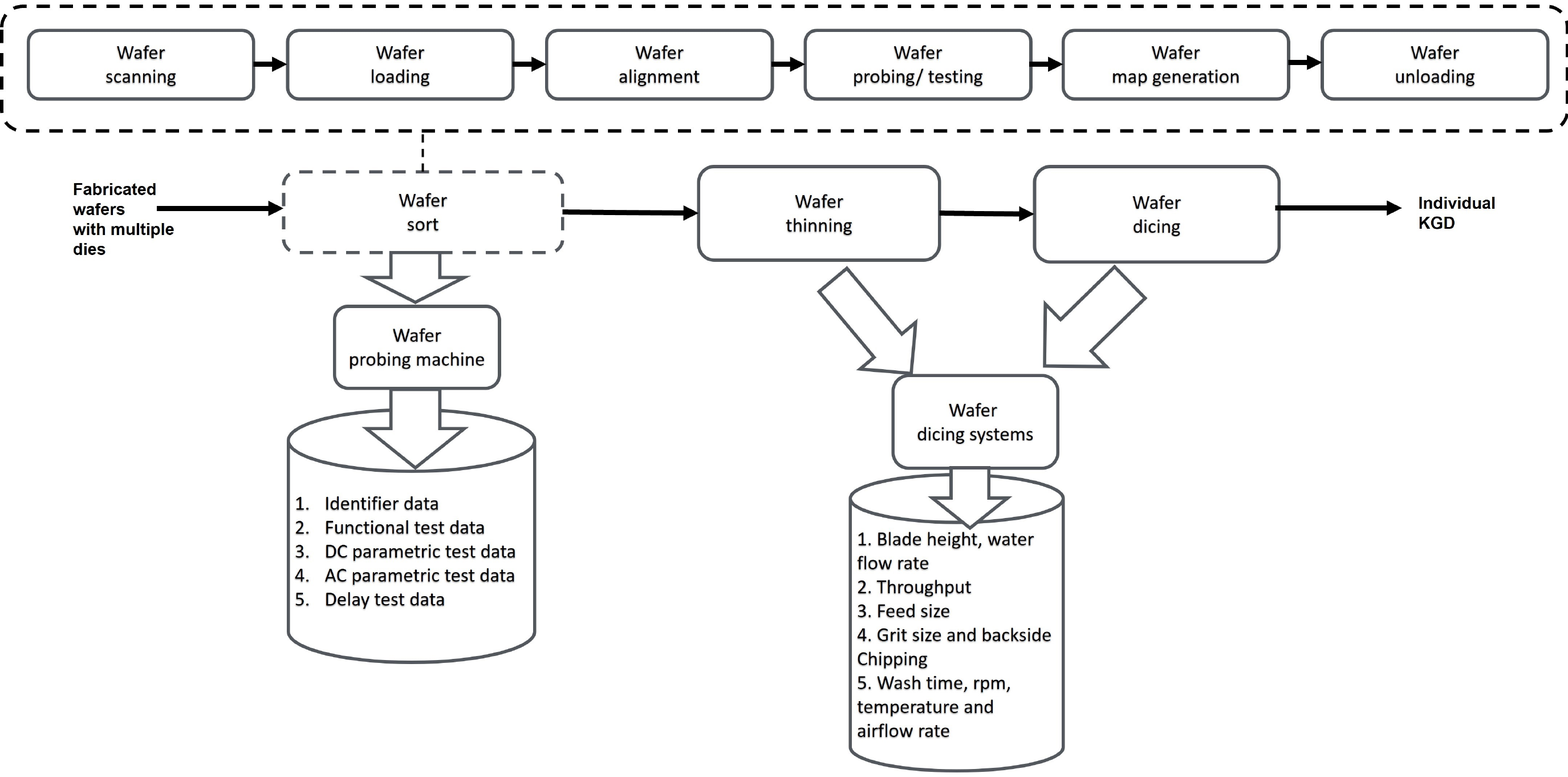}}%
\caption{Wafer testing process and possible data obtained. The dotted line around wafer sort signifies that it consists of sub-stages which are shown in the dotted line rectangle at the top.}
\label{fig: waf_srt}
\end{figure*}

The blank wafer is first thermally oxidized to form a layer of $SiO_{2}$ on top of it. 
This process is known as oxidation.
The oxidation step is carried out in oxidation ovens or furnaces which can tightly control the temperature and gas flow rate at which the oxidation reaction occurs. 
A real time monitoring of the oxide thickness and oxidation rate is also possible in modern systems\cite{Campbell2008fabrication}.
Then, different regions in the circuit such as active area, isolation, the gate, the p/n well, metal layers and contacts are patterned using photolithography through a series of chemical and mechanical process shown in Figure \ref{figure: fab}.

The dopant present in the source and drain areas is introduced through the bombarding the wafer surface by accelerated ions.

Deposition of metal and nitride layer on top of the wafer happens through either physical or chemical vapor deposition. 

In newer technology nodes, chemical vapor deposition (CVD) is used extensively to deposit nitride, oxide and even metal layers\cite{hampden1995chemical}. 
The CVD systems, similar to oxide furnaces, also provide the foundry with extensive real time data on the temperature, pressure, reactant flow rate, reaction time and growth diagnostics. 
The spin speed, applied torque, developed resist thickness and uniformity data, baking temperature, development time, spray pressure data are all available from the modern cluster equipment that can combinedly perform photoresist coating, developing and stripping.
The etching system carries out the etching of materials underlying the photoresist in either dry or wet etching process. 
Etching rate, trench slope, gas flow rate are example of data that can be obtained from an etching system.

The aforementioned data from these machinery is collected through built-in sensors. 
These are also known as in-line, online or in-situ data.
In addition to in-situ data, foundries also utilize a wide range of ex-situ tests to provide a stronger and more precise feedback of fabrication process parameters. 
Ex-situ tests, otherwise known as offline tests, are also generators of a large volume of data.
Some example of offline tests are as follows:
\begin{itemize}
    \item \textbf{C-V profiling:} Through the application of a DC voltage, the width of the space charge region in the junction area of a MOSFET may be manipulated. Using this principle, the C-V profiling test can determine the type of the dopant as well as measure the doping density.
    \begin{figure*}[h]
\centering
\subfloat{\includegraphics[width=7in]{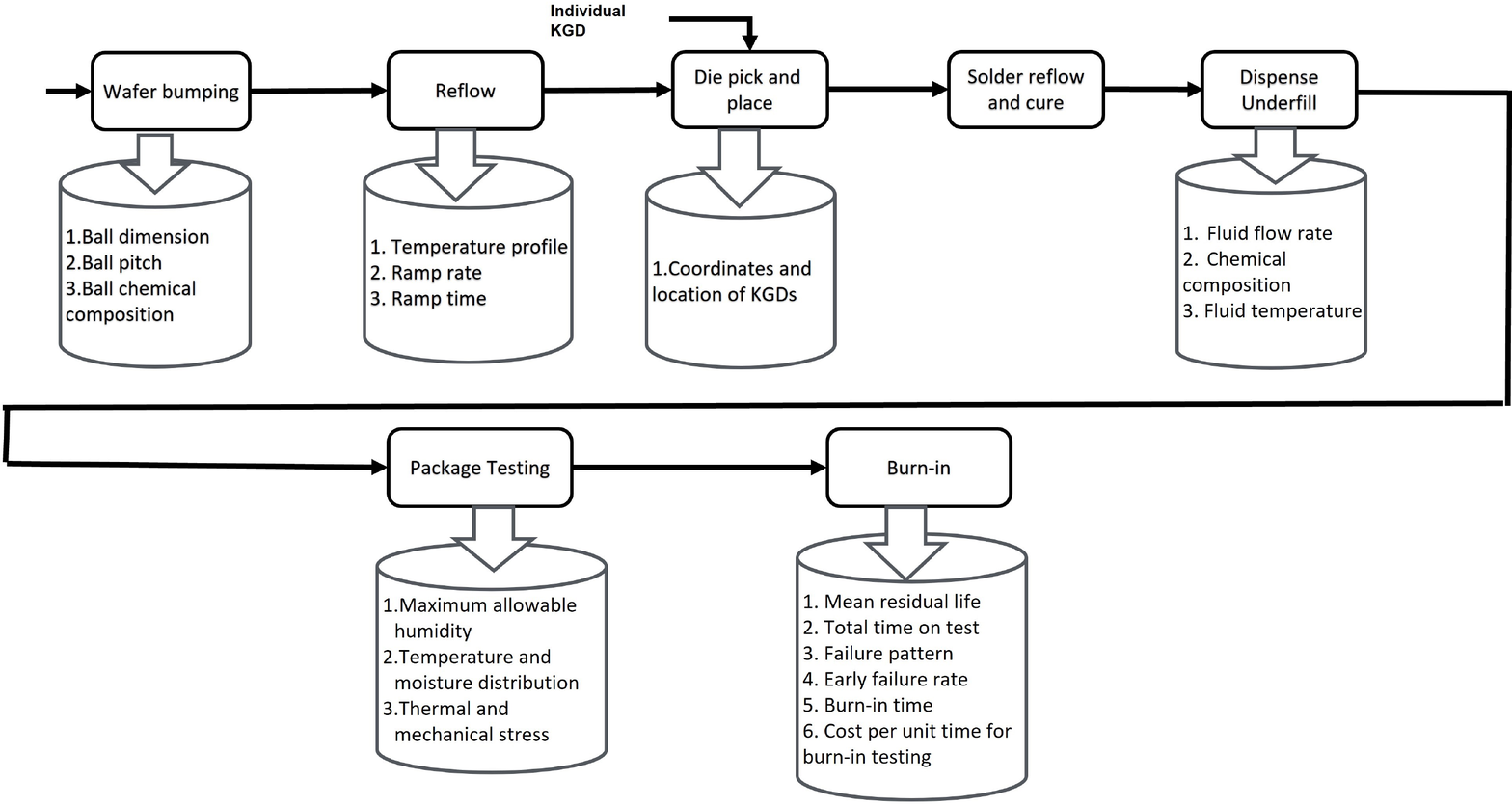}}%
\caption{Data available from assembly, packaging and reliability test flows. The packaging flow shown is for Flip Chip Ball Grid Array (FCBGA) packaging technology.}
\label{fig: asmbly,pkg,rel}
\end{figure*} 

    \item \textbf{Four-point probe:} A typical methodology for measuring semiconductor resistivity, linewidths and sheet resistance is the four-point probe method. It is most usually employed as an absolute measurement that does not rely on calibrated standards. Four point probe tests can be used to construct sheet resistance contour maps which in turn can be used to infer the doping density profile of the wafer. 
    \item \textbf{Thermal wave imaging:} The wafer is subjected to periodic heating stimuli. In the vicinity of the surface, the heating stimuli produces minute volume changes. These are detected with a laser by measuring the change in reflectance of the incident or pump laser\cite{smith1985ion}. The test data is represented in the form of a contour map.   
    \item \textbf{Microscopy:} Scanning capacitance microscopes (SCMs) or scanning spreading resistance microscopes (SSRMs) can be used to build a later doping profile of the sample \cite{neubauer1996two,vandervorst2001towards}. Atomic force microscopes (AFMs) and transmission electron microscopes (TEMs) can capture images up to nanometer resolution. These high resolution images can be used for a variety of purposes including critical dimension control, topography analysis, electrical potential measurement etc. 
\end{itemize}

\subsection{Post-silicon Packaging, Assembly and Test}
\label{subsec: package}
\subsubsection{Wafer Test}

After the end of BEOL in the wafers with multiple dies are loaded onto an automatic test equipment (ATE) called wafer probe station. 

A wafer prober is a highly sophisticated equipment that applies test patterns to check whether a given die on a wafer meets functional and parametric requirements based on which a chip is either accepted or rejected for packaging. 
The dies that are accepted and rejected together form a color coded wafer map which can be viewed by an operator in a computer. 
The entire wafer sort process is illustrated in Figure \ref{fig: waf_srt} along with the data obtained from these steps.

The wafer probing test itself applies test patterns in response to which the following data is gathered\cite{bushnell2004essentials}: 
\begin{itemize}
    \item [a.] \textbf{Functional test data:} No. of stuck-at, transistor open and short faults, the nets they occur in, fault and test coverage, no. of untestable of faults, automatic test pattern generation (ATPG) algorithm effectiveness.
    \item[b.] \textbf{Electrical parametric test data:} Various AC and DC parameters such as output drive and short current, contact resistance, input high and low voltages, terminal impedance and reactance. 
    \item[c.] \textbf{Delay test data:} The rise and fall times of transition, setup and hold times of sequential circuits are some of the data available from delay tests performed at wafer test step.
    \item[d.] \textbf{Test identifier data:} These data include device, lot and wafer ID, wafer flat position which is used for aligning, no. of wafers discarded after wafer probing. 
\end{itemize}
After the wafers have been probed, the dies that pass performance and functional requirements are sliced in wafer dicing systems. 

The data available from wafer dicing systems include the blade rpm, wash time, temperature, water flow rate etc. are also shown in Figure \ref{fig: waf_srt}.

\subsubsection{Packaging and Assembly}
\label{subsubsection: package}

Packaging refers to the process of encapsulating the known good dies (KGDs) in protective insulating material and attaching metal balls or pins to them so that they can be accessed from the outside. 

Assembly refers to the process of binding all of these different ICs and electronic components to a printed circuit board (PCB).
Rapid device scaling, growth in the number of I/O pins, necessity of access to DfT and DfD features, thermal, mechanical and economic considerations has meant that packaging technology has continually evolved over the past 60 years. 

Through the years, packaging technologies such as surface-mount technology (SMT), quad flat packaging (QFP), pin grid array (PGA) and ball grid array (BGA) have been used. 
Recently, 2.5D and 3D packaging technology have also been proposed. 
Depending on the particular technologies being used, the steps followed in the packaging process would be different. 
Figure \ref{figure: substage} shows the steps for BGA packaging, more specifically flip chip BGA (FCBGA) packaging.

After the bond pads have been opened, a metal 'bump' or ball is deposited on top of these pads. 
This process is known as bumping.
These bumps will form the bond between the substrate of the PCB and the die when the dies are 'flipped' to be conjoined.

The wafer is then diced and KGDs are picked and placed by an automatic machine to its appropriate place on the substrate ball side down. 

An epoxy type material is deposited by capillary action underneath to fill the space between the balls and the package. 
This step is known as underfilling. 

Underfill flow rate, chemical composition and fluid temperature are some examples of the data available at this stage.

\subsubsection{Package and Burn-in Testing}
\label{subsubsection: Burn}
Once the packaging and assembly steps are completed, the fabricated ICs are subjected to elaborate stress testing comprised of package and burn-in testing to evaluate their longevity under real world operating conditions.
Combined they are also sometimes known as reliability tests. 
Preconditioning, temperature cycling, thermal shock, temperature-humidity accelerated stress testing form a partial list of the series of stress tests that the chip is subjected to\cite{andrea2017}. 
Some of the data items available from this series of stress tests are shown in Figure \ref{fig: asmbly,pkg,rel}. 

\subsection{In-field Deployment Stage}
\label{subsection: deploy}

On-chip performance, voltage, temperature monitors monitor relevant circuit and software parameters and collectively form a report on the health and performance of the device.
The availability of the tests listed in the following therefore, depends largely on the on-chip sensors, DfT facilities, and the interfacing software for a particular chip.
Different vendors also enhance the existing standards to offer additional debugging and testing features into their chip and as such, the data available from these tests would largely depend on the specific vendor and type of the chip.
\begin{enumerate}
    \item Built-in Self Test (BIST): BIST is used to periodically test the circuit subsystems and their operation\cite{nourani2008low}. Its main purpose is to verify whether different components are working properly and in some cases, apply appropriate countermeasures. Two types of BIST are widely used: Logic (LBIST) and Memory (MBIST). LBIST generates input patterns for internal scan chains using a pseudo-random pattern generator such as linear feedback shift register.
      MBIST is used for detecting memory defects and in some cases, repair those defects.
    \item Joint Test Action Group (JTAG) debugging: JTAG is a standard to access the boundary scan DfT features in a chip to verify its functionality\cite{tehranipour2003testing}. Although originally conceived as a means of overcoming limitations of bed-of-nails fixtures of testing PCBs after manufacturing, today it is used for diagnosis of failing interconnects, memories, and testing functionality of ICs. Often a boundary scan description language (BSDL) specification of existing JTAG features on a chip are provided by vendors to customers. This ensures that customers can have a useful manual on what test features are present in their device and how to use them.
    \item Hardware Performance Counters (HPC): HPCs are special purpose registers provided in a chip that stores various performance related activities in the device. These statistics can usually be accessed by an operating system (e.g., in Linux these may be accessed by the \textit{perf} instruction) or special purpose software for the purpose. 
A list of data available from these tests may be found in Table \ref{table: in-field}. 
\begin{table}[!t]
\caption{Data Available from In-field}
\label{table: in-field}
\centering

\begin{tabular}{p{0.15\textwidth} p{0.28\textwidth}}
\hline
\textbf{In-field source of data} & \textbf{Available Data}\\
\hline
BIST & Different subsystems in the device such as UART, memories, 
LED system etc. working 
or not. Certain vendors and chips may offer additional functionality to show coverage of stuck-at faults and memory faults tested.
\\
\hline
JTAG debugging & Interconnect open and 
shorts and associated nets/
units, existence of stuck-at, crosstalk faults, device ID, mon-testable nets and 
coverage statistics, real-time program counter.
\\
\hline
BSDL description & JTAG instructions and 
available registers, signal mapping, package
information, type of boundary cell
available for signals.
\\
\hline
Hardware performance counters & Cache references, branch misses, bus cycles, cycles, cache misses, CPU-cycles, L1-dcache and L1-icache loads, stores and load-store misses, LLC load, stores, load-store misses.
\\
\hline
\end{tabular}
\end{table}
\end{enumerate}

\section{Existing Hardware Attack Vectors: Data and Security Perspectives for DT}
\label{sec: HAV}
 
To keep our discussion focused, we describe three hardware security threats associated with the scenarios we mentioned earlier in section \ref{sec: motivation}, namely: hardware trojans, counterfeits and information leakage. 
In this section, we also highlight the data items that have a correlation with these attack vectors.

\begin{sidewaysfigure*}[htbp]
\centering
\subfloat{\includegraphics[width=9.25in]{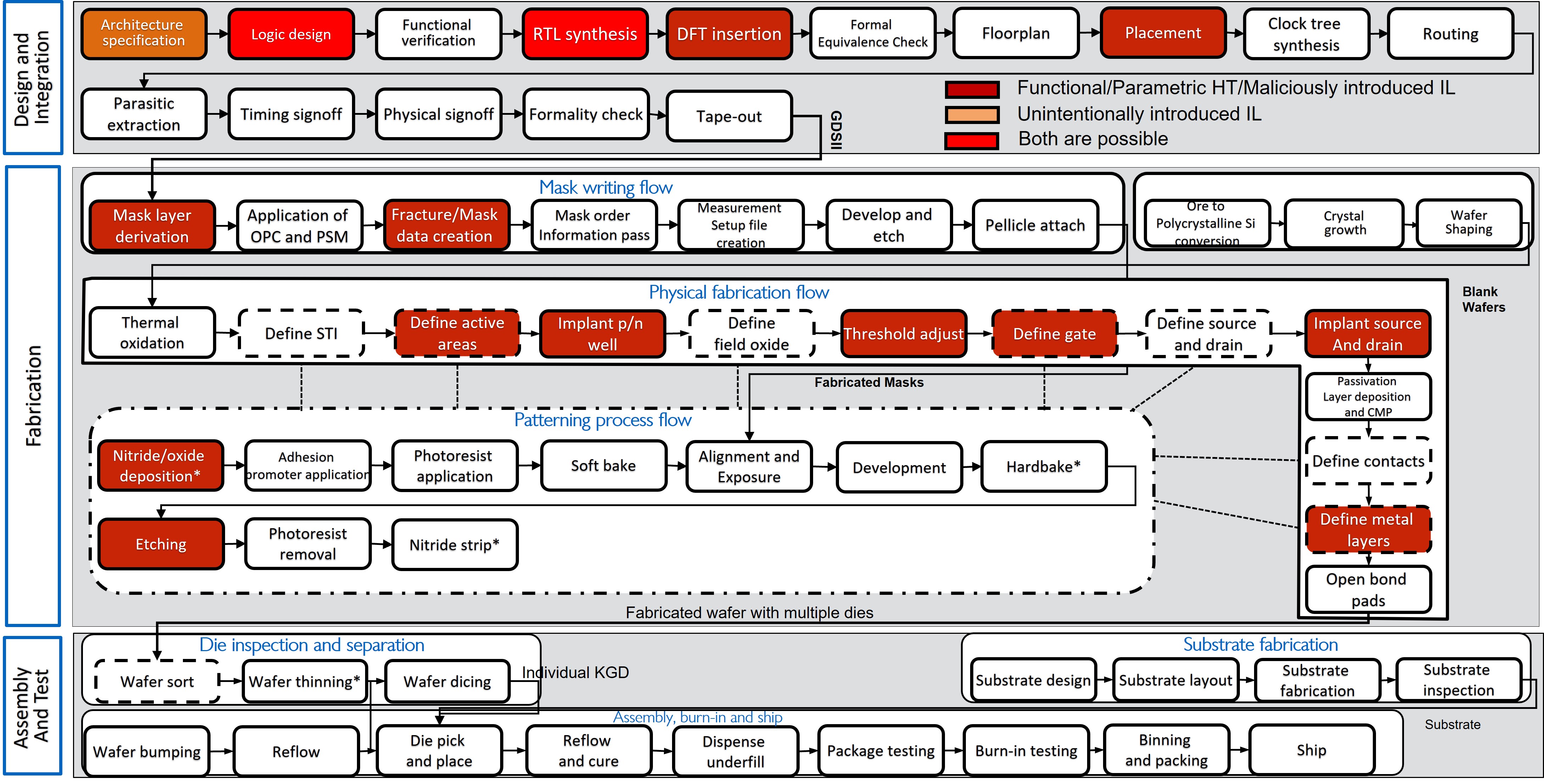}}%
\caption{Example lifecycle substages where a hardware trojan inserted or information leakage vulnerability may be introduced unintentionally. Hardware trojan associated stages are marked in red whereas information leakage stages are marked in orange. It should be noted that information leakage may be caused intentionally by both functional and parametric trojans. So, the red marked boxes may be considered as source lifecyle stages of intentional information leakage.}
\label{fig: threats}
\end{sidewaysfigure*}

\subsection{Attack vectors}
\subsubsection{Hardware Trojan (HT)}
\label{subsec: Hardware Trojan}
A Hardware Trojan (HT) is a term that refers to a malicious alteration of a circuit.
The Trojan may take control of, alter, or obstruct the underlying computing device's components and communications. 
Trojan trigger refers to the specific condition or the circuit that leads to the dormant trojan being activated.
Trojan payload refers to the functionality that a trojan achieves once it has been activated.
Trojan payload may either leak sensitive information to an externally observable port, cause denial of service, degrade performance, cause accelerated aging, or change the functionality entirely.
There are a variety of ways in which hardware trojan may be categorized. 
A detailed trojan taxonomy is provided in \cite{tehranipoor2010survey}. One important distinction we want to note here is between functional and parametric trojans. 
Functional trojans are trojans that inherently change the functionality of the circuit whereas parametric trojans manipulate certain electrical or physical parameters of the device to cause performance degradation.

In Figure \ref{fig: threats}, the red marked boxes show the lifecycle stages where either a functional or parametric trojans may be inserted.
In the design phase, functional hardware trojans may be inserted through acquired 3PIPs. 
In the logic design, RTL to gate level synthesis and placement phases these IPs are referred to as soft, firm and hard IPs respectively.
DfT features, when outsourced to a third party test vendor, may also be a source of trojans.
The trojans inserted in the design phase are almost all exclusively functional trojans.
Functional trojans may also be introduced by the untrusted foundry.
In that case, functional trojans would require the modification of the GDSII file provided by the design house which translates to a manipulation of the mask writing data in the mask layer derivation and mask data creation phases. 
Due to the need of adding additional logic into the circuit, for this type of trojan additional active areas need to be included.

Parametric HTs may be inserted by manipulating fabrication recipes across a wide range of steps in the physical fabrication flow shown in Figure \ref{fig: threats}.
For example, doping dosage, oxide thickness\cite{shiyanovskii2010process}, threshold voltage\cite{patil2017manufacturer} may be manipulated to cause accelerated aging or parametric failure of the device. 
Critical dimension change of the gate and channel lengths by changing the etching depth can also cause such aging.

Trojan detection is challenging even for state-of-the-art detection techniques.
To begin with, the intrinsic opaqueness of integrated circuit internals makes it difficult to identify manipulated components; typical parametric IC testing procedures are often ineffective due to limited coverage during testing.
Even if a testing method could be devised that reach extremely wide coverage, hardware trojans can be sequentially activated meaning only a very specific sequence events can trigger it. 
Formal verification methods often fail due to state explosion problems in such cases.
Destructive tests and IC reverse engineering techniques are time consuming and costly.
When technology scales to the boundaries of device physics and mask imprecisions, a chip's properties become nondeterministic, making the difference between what is a device affected with merely process variation and a device infested with Trojans difficult to surmise. 
Finally, the layout of the design may have `empty' spaces that serve as HT insertion spots for a malicious foundry.

\subsubsection{Counterfeits}
Due to the complex globally distributed horizontal nature of the electronics supply chain, it is difficult to trace the authenticity of each component that goes into an electronic system.
The most frequent hazard associated with an untrustworthy electronics supply chain is the availability of various forms of counterfeit devices.
Counterfeits can be of the following types:

\begin{itemize}
    \item \textbf{Recycled:} The recycled electronic components are recovered from used PCBs that are disposed of as e-waste, repackaged, and resold in the market as brand new components. Despite the fact that such devices and systems may still be functional, there are performance and life expectancy difficulties associated with them because of aging process and various adverse effects resulting from exposure to chemicals during the recycling process. Recycling therefore, is an end-of-life issue from the viewpoint of provenance.
    \item \textbf{Remarked:} Electronic components that have had the labeling on their package or the die replaced with falsified information are known as remarked chips. New electronic equipment might also be intentionally mislabeled with a higher standard by the untrusted foundry or other actors in the supply chain. For example, a chip may be designated as industrial or defense grade despite only meeting the requirements of a commercial grade one.
    \item \textbf{Overproduced:} Untrustworthy foundries, assembly plants, and test facilities that have access to the original design may responsible for overproduction. These parties may be able to fabricate more chips or systems than the number specified in the contract and resell them without permission. Overproduction thus originates from the fabrication phase of the lifecycle.
    \item \textbf{Defective and/or Out-of-specification:} Failure to comply with functional or parametric standards or grades (e.g., commercial, industrial, and military) results in the rejection of the device during fabrication and testing phases. However, the untrusted foundry or testing facility may ship defective or out-of-spec components into the market as genuine integrated circuits or systems without the knowledge of the design house.
    \item \textbf{IC cloning and IP theft:} A cloning attack can be carried out by any untrustworthy party in the electronics supply chain. Clones are direct copies of the original design that are created without the consent of the original component manufacturer (OCM). Cloning may be accomplished by reverse engineering an IC or system that has been purchased from the market. IP theft refers to the stealing of intellectual property design components or tools and selling them to other parties without compensating the original owner. This may include things like the HDL code of IP cores. Cloning may happen at any time during post silicon phases where as IP theft is mostly a pre silicon issue.
\end{itemize}
Due to the overarching nature of the origin of these counterfeit types insofar as lifecycle substages are concerned, they have not been explicitly shown in Figure \ref{fig: threats}. 
Various approaches have been proposed in literature to combat counterfeits. 
For preventing the shipping of defective, out-of-spec and overproduced chips hardware metering approaches such as Secure Split Test (SST) have been proposed\cite{contreras2013secure}. 
Although these methods are effective, due to the requirement of new industrial practices for successful realization, these have not yet been fully integrated into the traditional lifecycle and many design house still have to almost blindly trust the foundry to get their chips into the market.
Aging based statistical analysis parametric fingerprints are used for detection of recycled ICs\cite{huang2012parametric,zhang2012path} but they may suffer from reduced accuracy due to process variations.

\subsubsection{Information Leakage (IL)}
Information leakage refers to the breach of integrity and/or confidentiality requirements in the security policies of a device.
Confidentiality requirement violation results in unauthorized parties being privy to sensitive assets on the device while integrity violation results in such parties being able to modify these assets.
Information may be leaked through primary debug or test access ports due to unintentional mistakes made in the design phase or architecture specification stage by the designers or due to the insertion of a malicious hardware trojan\cite{nahiyan2017hardware}.
They can also be leaked unintentionally through side channels such as timing, power, acoustic etc.
Inserted hardware trojans can be responsible for leaking information through observable points in the circuit. 
In summary, there are two types of IL: maliciously introduced through HTs and unintentionally introduced.
Maliciously introduced IL sources in lifecycle stages are highlighted in red while unintentional ones are highlighted in orange.
In the logic design and RTL to gate level synthesis steps both types of IL may be introduced.
The IL vulnerability unintentionally introduced in RTL to gate level synthesis step is mainly due to CAD tools that do not take security concerns into consideration. 

The challenge in detecting intentional IL caused by the insertion of a hardware trojan are the same as they are for hardware trojans.
Unintentional IL can be detected by formal verification methods\cite{rajendran2016formal} and information flow tracking (IFT) methods\cite{hu2014gate}. 
Formal verification methods are entirely reliant on the expertise of the verification engineer and the capability of the verification software.
If formal assertions are not written properly, they might throw false positives. 
Model checking software used for formal verification also have the state explosion problem as the design can be `unrolled' only to a limited number of cycles.

In addition to the challenges outlined above, hardware attacks are evolving and new threats are proposed in literature that circumvent traditional and literature proposed detection schemes.
For example, there are always new trojans being designed by researchers and malicious actors alike that defeat existing traditional verification methods as well as the ones proposed in literature\cite{liu2016silicon, kison2019security}. 
So it is entirely feasible that a hardware trojan may go unnoticed until it is triggered in the field or in a more advantageous situation, detected at a later testing stage than the one it was inserted in. 
However the case may be, once it has been detected, it is impossible for almost all existing hardware trojan detection techniques to inform the defender on where the trojan was inserted.
For instance, if a trojan is detected at the formal equivalence check step, the designer can not infer, merely from the result of the detection test, whether it was a soft or firm IP or the test vendor that introduced the trojan.
Similar arguments may be made for all hardware attack vectors including IL and counterfeits discussed in this section.
This is where the proposed DT architecture adds new dimensionality to the hardware security threat analysis and defense.

\begin{table*}[!h]
\caption{Relationship between Available Data and Hardware Security Threats}
\label{relation_table}
\centering

\begin{tabular}{p{0.22\textwidth} p{0.2\textwidth} p{0.22\textwidth} p{0.22\textwidth}}
\hline
\textbf{Data item} & \textbf{Related security vulnerability}& \textbf{Available from (Stage)} & \textbf{Available from (Equipment/test/software)} \\
\hline
Diffracted intensity & Hardware trojan, maliciously introduced information leakage & Mask writing & Lithography simulation software\\
\hline
Branching probability, Relative branching probability, Controllability Index & Hardware trojan, maliciously introduced information leakage & Logic Design & RTL parser/ static HDL code analyzer\\
\hline
Code, functional, and toggle coverage & Hardware trojan, maliciously introduced information leakage & Formal verification &  Cadence JasperGold® \\
\hline
Pattern area density, shot time, dosage of radiation & Hardware trojan, maliciously introduced information leakage & Mask writing & E-beam mask writer systems\\
\hline
Out\_flipflop\_x, in\_flipflop\_x, in\_nearest\_pin etc. & Hardware trojan, maliciously introduced information leakage & RTL to gate level synthesis & GLN parser\\
\hline
OPC program runtime and filesize & Hardware trojan, information leakage & OPC correction & OPC software such as Cadence® Process Proximity Compensation (PPC)\\

\hline
Doping density & Hardware trojan, maliciously introduced information leakage & Ion implantation & Ion implantation systems, Four point probe, Thermal wave imaging, C-V profiling\\

\hline
Gate and oxide dimension & Hardware trojan, maliciously introduced information leakage & Gate definition, oxidation & Four point probe, AFM/TEM, oxidation furnace\\
\hline
Etching depth & Hardware trojan, maliciously introduced information leakage & Etching, post-fabrication & Etching systems, AFM/TEM\\
\hline
SCOAP controllability and observability & Hardware trojan & RTL to gate level synthesis & Synopsys TetraMAX®\\
\hline
Contour map & Hardware trojan, maliciously introduced information leakage & Any time after ion-implantation & Modulated photoreflectance, four-point probe, C-V profiling\\
\hline
Lead plating, ball chemical composition & Recycled, remarked, cloned & Wire bonding & Energy dispersive microscopy \\
\hline
Texture of package &  Recycled, cloned & Package test & Any photograph taken of golden chip available in the market or after fabrication\\  
\hline
Bond wire, ball, pin dimension and count & Recycled, defective & Wire bonding, package test & Visual inspection or microscope imaging \\
\hline
Invalid markings on the package (e.g., lot identification code, CAGE code, pin orientation marking) &  Remarked & Any time after burn-in test & Any photograph taken \\
\hline
Lead, pin, or ball straightness, pitch, alignment & Recycled & Wire bonding, flip chip attach & Visual inspection, microscope image \\
\hline
 No. of good and functional parts tested & Out-of-spec/defective, overproduced & Wafer sort & STDF database (Master results record)\\
\hline
Bin no., number of parts in the bin & Out-of-spec/defective & Wafer sort, various offline and inline test performed in the manufacturing floor & STDF database (Hardware bin record)\\
\hline
Wafer ID, No. of good parts tested per wafer & Defective/out-of-spec, overproduced & Wafer sort & STDF database (Wafer results record)\\
\hline
Early failure rate & Recycled & Burn-in testing & Wafer prober or burn-in tester\\
\hline
Curve trace & Different types of counterfeits & Wafer sort & Wafer prober\\
\hline
\end{tabular}
\end{table*}

\subsection{Data and Security Perspectives}
\label{subsection: security}

\subsubsection{Hardware Trojan \& Information Leakage}
Maliciously introduced information leakage may be caused by trojan insertion and as such any data item that is related to hardware trojan insertion is also related to maliciously introduced information leakage.
We list some data items that are related to HT and IL threats which are already available from traditional flows of the semiconductor lifecycle.
\begin{itemize}
    \item Pattern density refers to the number and width of features that need be transferred from a mask to the wafer in unit area of the mask\cite{rizvi2018handbook,takahashi2000proximity}.
    If functional trojans are inserted, the pattern area density available from mask writing tools might be different from what was nominal  for the layout delivered with the GDSII file as heretofore absent features need to be added to the mask so that it translates to the new malicious logic added to the circuit.
    \item Electron Beam Lithography (EBL) systems use electron beams to etch pattern onto the mask according the output of mask layer creation software. 
    The shot time of incident electron beams is proportional to pattern density which means that if hardware trojans are inserted, shot time may be changed as well.
    The governing equation is $T=\frac{T_0}{1+2\alpha\eta}$ where $T_0$ is the shot time at zero pattern density and T is the shot time at $\alpha$ pattern density.
    \item Electron Beam Proximity errors are encountered in mask writing when neighboring features are too adjacent. They need to be compensated for by dosage correction. This dosage correction is inversely proportional to pattern density which in turn is related to hardware trojan insertion.
    \item If functional hardware trojans are inserted, it can be reasonably deduced that to add the necessary logic into the circuit, more material will need to be etched at most of the fabrication steps. The more material that needs to be etched, the more the etching plasma will be depleted, and hence reduce the etch rate.
    This is known as the etch-loading effect and it is typically more pronounced for dry etching. 
    This data is usually available from the etching system used in photolithography process.
    \item When hardware trojans are inserted by a malicious foundry they are typically inserted in the empty spaces in the layout. This increases the pattern density as mentioned before. Additionally, as the features increase in adjacency we can reasonably expect that the foundry must compensate for these additional features by adding more OPC features such as vertices and line segments. This should increase the file size and runtime of the OPC program.  
    \item Many hardware trojans are triggered by rare signals present in the circuit. The controllability and observability of nodes can be measured through SCOAP measure which is available with most modern commercial synthesis tools.
    \item The slope of an etched slope is inversely proportional to the density of features. So, addition of functional trojan features may impact the slope of the features which might be visually inspected by high resolution microscopy images.

    \item The intensity profile of diffracted rays through the mask are effected if patterns are too close by. This data is available from lithography simulation software
    \item Critical dimension manipulation can be checked for through offline test methodologies or through high resolution microscopy imaging. Critical dimension data such as oxide thickness, gate dimension are related to parametric trojans. These data are also available from the respective production equipment in which these steps are performed.
    \item Recipe changes to insert parametric trojans such as doping dosage change, etching depth change can be traced from different offline tests as well as inline sensors in the production equipment. The contour map generated from offline tests such as thermal wave imaging and four point probe are indicative of the doping ion used and doping dosage. So any change in doping dosage may be traced back to the results of these tests.
    \item Branching probability metrics proposed in \cite{choo2020register}, which measure the likelihood of certain branches in RTL code being taken, can be extracted from HDL description of the circuit. These features are correlated with hardware trojans activated on rare condition triggers. 
    \item Authors identified 11 most important features extracted from the netlist of a circuit related to hardware trojans in \cite{hasegawa2017trojan}. These include features such as how many logic levels away a net is from the primary output/input, the number of FFs a certain logic level away from that net etc.
    \item For unintentional information leakage verification, formal verification methods can be useful. Although care must be taken to consider the possibility of false positives and the limited capability of the verification tool.
\end{itemize}

\begin{figure*}[!t]
\centering
\subfloat{\includegraphics[width=7in]{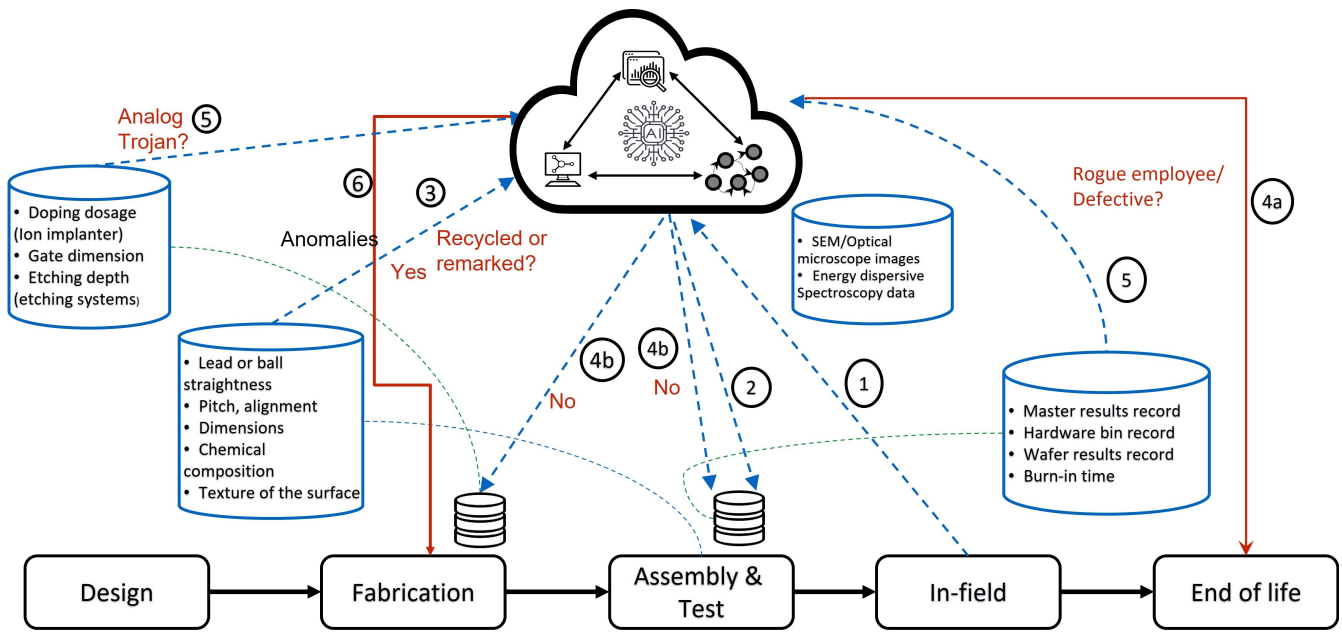}}%
\caption{An example of how various data and information from different lifecycle stages can be used to trace and reason for an observed anomaly, as described in scenario 1.}
\label{fig: DT_data}
\end{figure*}

\subsubsection{Counterfeits}
Following are some of the data that can be analyzed to relate to different types of counterfeits:
\begin{itemize}
    \item Package markings are carefully placed by OCMs. Package markings may include things like country of origin, lot identification number, pin orientation marking etc. A deviation from OCM provided specifications indicates remarking of the chip.
    \item The pitch, alignment of pins, balls and leads are carefully selected by OCM. Deviation indicates potential recycling. The dimensions of leads are also carefully selected and monitored and as such deviation may indicate a recycled/remarked chip.
    \item Leads, pins or balls of a chip may be reworked during recycling resulting in a different material than that of the authentic chip. A cloned chip’s lead might not also have the same material composition as the original. This material composition can be verified from tests like energy dispersive microscopy.
    \item Crudely recycled or defective chips may have damaged bond wires. The length, shape and tolerable stress of these wires may be found from the wire bonding stage. Wire bonding stage is common for DIP and QFP packaging technologies. This stage does not exist for BGA packages which is why it is not shown in Figure \ref{figure: substage}.

    \item The standard test data format (STDF) is a widely used specification in the semiconductor industry\cite{STDF}. Many ATEs used in the industry upload their test data to the database maintained by the foundry or the test facility in this format. It has also evolved as the de facto standard for organizing and collecting other test data from the fabrication and test facility. The STDF test specification lists 25 different types of record that are catalogued in a STDF file. Among these wafer result and master result records are of interest in dealing with overproduced, defective and out-of-spec chips. Combined these keep a track record of the no. of parts that were tested to be of acceptable quality in each wafer. As such, these might be verified to trace whether yields were falsified. Furthermore, hardware bin records are kept to list which chips were placed in which bin in the testing facility. So, if a rogue employee is shipping defective chips out of the foundry, this record may be analyzed to find evidence of unaccounted dies and chips. STDF also comes with audit trail record which keeps a detailed history of each modification made to the file. This can help in dealing with insiders that might be falsifying the test data.
    \item Curve tracing and early failure rate, available from parametric testing and burn-in testing, respectively can be performed to detect counterfeit ICs\cite{guin2013anti}.
\end{itemize}

A summary of these relationships are listed in table \ref{relation_table}.
These relationships establish that if anomalies in relevant data items are found, then it could be reasoned automatically \textit{where} an attack took place. 
Figure \ref{fig: DT_data} illustrates how some of these dataset may be used to trace the root cause of the observed accelerated failure of chips described previously. For example, we start with some in-field data or report collected for the chip for an operation-of-interest, i.e., accelerated failure in this example . 
After a report of the observed accelerated failure is uploaded to the DT, DT would reason that one of the three hardware vulnerabilities mentioned earlier is the probable source of it . 
Then, the DT would require images obtained from microscopy of the device and/or a simple high resolution photograph(shown with \textcircled{1} in the figure). 
Data from more involved tests such as energy dispersive spectroscopy may be uploaded to the DT as well. 
At this point, from historical database of assembly and packaging stage, the DT crosschecks the uploaded data with data items that have a relationship with recycled and remarked chips (e.g., lead, ball dimensions, count, chemical composition, texture of the surface of the chip)(\textcircled{2}-\textcircled{3}). 
If the uploaded data do not match with previously stored specifications, then it is likely a recycled or remarked chip(\textcircled{4a}).
However, if anomaly can not be found then other explanations must be explored. 
As such, the DT would look at data associated with process variation based trojan related data (e.g., gate and oxide dimension, etching depth, doping dosage etc.) and shipping of defective/out-of-spec chip related data (e.g., master results and wafer results record in the STDF database) to determine if there is an observed anomaly(\textcircled{4b}-\textcircled{6}).
There are some pertinent issues to be mindful of. 
Anomaly detection algorithms for any of these data items may not be robust and thus there is a significant element of uncertainty in assigning a deterministic cause to any of these underlying possible explanations.
Uncertainty may also arise from the fact that collected data may be inadequate or too noisy to reach a conclusive decision.
Therefore, the DT for secure semiconductor lifecycle would require an AI algorithm that can model this uncertainty and reason accordingly. 
The AI algorithms that may be used for this purpose is explained in more detail in section \ref{sec: DT structure}. 

\begin{figure*}[!t]
\centering
\subfloat{\includegraphics[width=6in]{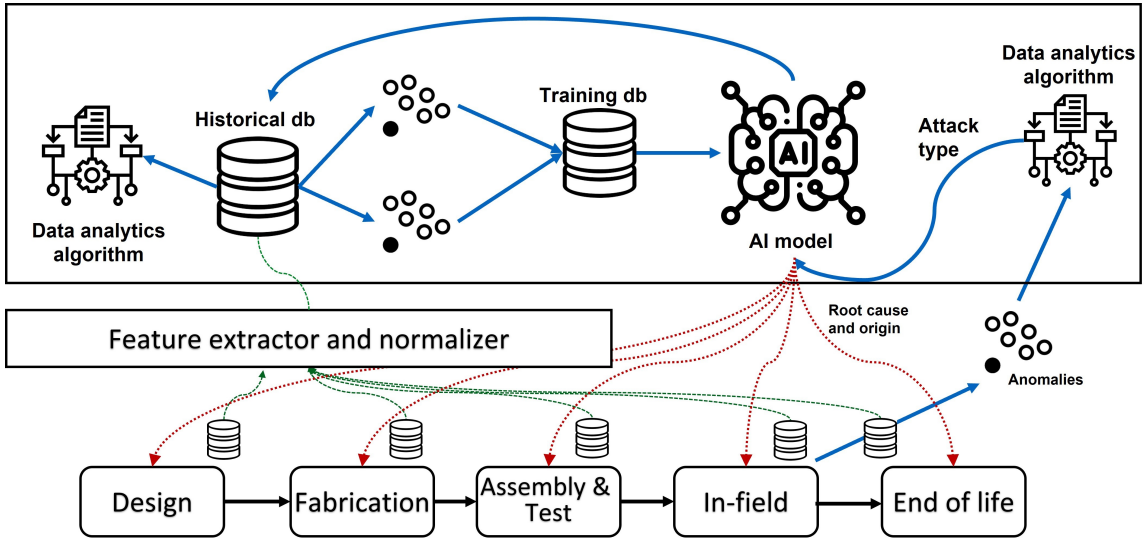}}%
\caption{A more in-depth view of proposed AI based DT framework. For the scenario 1, the root cause analysis starts with observation of an anomalous behavior in the field. This anomalous data is uploaded to the DT which runs its own analytics to figure out the possible attack types. This information is passed onto an AI model which consults a historical database. Based on the plausible attack vectors, the DT consults a subset of the historical database i.e., fabrication, assembly and in-field databases. Using further analytics, the DT finds traces of anomaly in the uploaded data and forms a training database with which the AI model is trained. The AI model then performs statistical inference on the database to find the root cause.}
\label{fig: DT_full}
\end{figure*}
\section{Digital Twin Structure and Modeling for Secure Semiconductor Lifecycle}

We illustrated in section \ref{subsection: security} how different data collected across different stages of the lifecycle may be utilized by a DT to provide traceability to observed anomalous behavior in the lifecycle. 
However, the view of the framework in Figure \ref{fig: DT_data} is not complete since the full functionality and underlying algorithms were opaque. 
We present the complete DT structure in Figure \ref{fig: DT_full}. 
The DT is driven by a collection of AI and data analytics algorithms.
There are four distinct components of the proposed framework:
\begin{itemize}
    \item [i.] The initial anomaly discovery is driven by a data analytics algorithm that uses verification, testing or sensor data to find violations in the specified security policies of the device. This component is shown at the top right side of Figure \ref{fig: DT_full}.
    \item [ii.] Feature extractors are simulation tools, scripts, emulators, EDA tools, parsers or scripts that extract and normalize features in context of the threat model and core AI algorithm.
    \item [iii.] Anomaly detectors find the threshold of relevant features to label instances within the database as anomalous. We argue that if a hardware attack has taken place one or possibly more of the features that have a correlation with the threats at hand would contain traces of anomaly. The anomaly detectors are data analytics algorithm (e.g., statistical models, time series analyzers) or even ML or DL models that can find said anomalies in the extracted features.
    \item[iv.] Using these evidences of anomaly, the core AI model will infer the lifecycle stage where the problem originated from. We note here that not all relevant features will contain conclusive evidence of anomaly in them. Consequently, we propose to use an AI algorithm to infer the probable cause in absence of complete consensus of extracted anomalies.
\end{itemize}
\label{sec: DT structure}

For the scenario 1 described earlier, the analysis commences with the observation of an anomalous behavior. 
In this case, the observed anomaly is a simple one i.e., accelerated failure of the chip.
Once the possible causes have been identified, the DT would consult its historical database. 
For scenario 1, the relevant databases are that of fabrication and assembly \& test stages. 
The DT would now require another set of auxiliary data analytics algorithm to find traces of anomaly in these databases. 
Once these anomalies have been found, these can be incorporated as evidence or knowledge base for training the central AI algorithm which is responsible for root cause analysis.
For the core AI algorithm we are proposing statistical relational learning in this paper although other approaches with similar capabilities would also suffice.
\subsection{Backward Trust Analysis}
We define the first functionality of the proposed DT for security assurance in semiconductor lifecycle as Backward Trust Analysis.
It has the following three components:
\begin{itemize}
    \item When an anomalous behavior is suspected in the performance of a chip, the DT should be able to analyze the uploaded data to confirm or deny that suspicion.
    \item If the suspicion is confirmed then the DT should also be able to identify what type of attack might the device be under. This step would require incorporating domain knowledge as well as data driven processes.
     \begin{figure*}[h]
\centering
\subfloat[Simple Bayesian Network]{\includegraphics[width=0.25\textwidth]{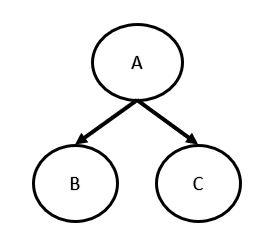}%
\label{fig: simple}}
\hfil
\subfloat[Possible formulation of a Bayesian Network describing causal relationships for scenario 1]{\includegraphics[width=0.9\textwidth]{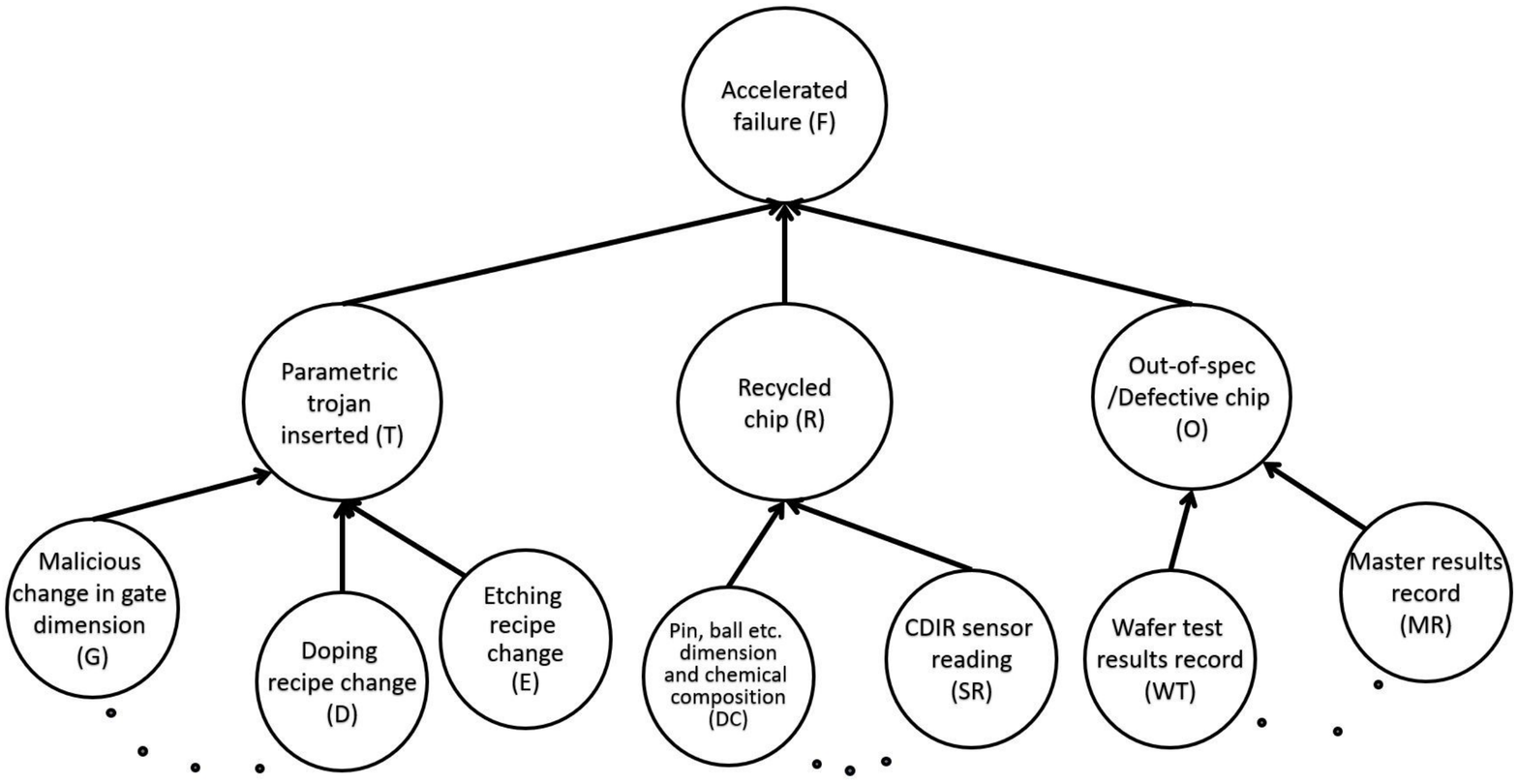}%
\label{fig: scenario1}}

\label{fig:BN_s1}

\caption{Bayesian Networks for root cause analysis.}
\end{figure*}

    \item Once possible causes have been identified, the DT should be able to analyze historical data to find traces of anomaly in relevant data items. Using these evidences of anomaly, the DT should be able to assign a \textit{most probable} cause to the observed behavior, determine where the attack took place and who the attacker was.
\end{itemize}
It is important to note that the data available from different stages of the lifecycle are not i.i.d (independent and identically distributed).
Traditional machine learning (ML) and deep learning (DL) algorithms are not suited for causal reasoning that is required for carrying out backward trust analysis as all of them implicitly carry the i.i.d assumption\cite{koller2007introduction}.
Consequently, we propose to use statistical relational learning (SRL) models to reason by combining probability theory, machine learning and mathematical logic. 
We discuss three possible SRL models, namely Bayesian Networks (BN), Hidden Markov Models (HMM) and Markov Logic Networks (MLN).

\subsection{Statistical Relational Learning Models for DT}
\label{subsection: srl}
\subsubsection{Bayesian Networks}
\label{subsubsection: BN}

As a probabilistic graphic model (PGM), BNs make use of Bayesian inference for probability computations. They describe conditional dependency between random variables as edges in a directed acyclic graph (DAG). 
Bayesian networks attempt to model conditional reliance and, by extension, causality between random variable through describing the joint probability distribution of constituent random variables. 
More specifically, a Bayesian Network $B(G,\Theta)$ over the set of random variables $\textbf{X}=(X_1, X_2, ..., X_N)$ has two constituents:
\begin{enumerate}
    \item A graph $G$ consisting of nodes representing the set of random variables and edges representing causal relationships among these variables.
    \item A set of conditional probability distribution $\Theta$ associated with each node of the graph.
\end{enumerate}

Figure \ref{fig: simple} shows a simple Bayesian Network which signifies that there is a causal relationship between events A and B or C. 
Event A can cause either event B or event C to happen which is signified by an `edge' pointing from A to B and C.
In BN terminology, A is the parent node whereas B and C are child nodes.
Starting from any node in the graph and following along the edges, one cannot end up in the same node that they started from. 
This is why it is called an acyclic graph.
The joint probability density function for any node in a BN is given by the formula:
\begin{equation}
    P(X)=\prod_{i=1}^{N} P(X_i | parents( X_i))
\end{equation}

\paragraph{Inference in Bayesian Networks}
Inference in BN refers to the process of finding the probability of any given node when conditional probability distribution of all other nodes are known. 
Inference in BNs can take two forms: the first is a straightforward calculation of the joint probability of a certain value assignment for each variable (or subset) in the network\cite{guo2002survey}. 
Because we already have a factorized representation of the joint distribution, we can simply evaluate it using the specified conditional probabilities. 
If we are only interested in a subset of variables, we must eliminate those that are irrelevant. 
This task is known as \textit{belief revision}.
The second one, is a calculation of the probability of a node $X$ given some observed values of evidence nodes $E$ i.e. $P(X|E)$. 
This type of inference is known as \textit{belief updating} or \textit{belief propagation}.
More formally, the task of belief updating may be represented by the following formula:
\begin{equation}
    P(X|E)=\sum_{\forall y \epsilon Y} P(x,e,y)
\end{equation}
where Y is the set of random variables that do not appear in either x or e.

\paragraph{Learning in Bayesian Networks}
The inference in BNs require the prior knowledge of the conditional probability table (CPT) of each node. 
CPTs of each node can be learnt in one of three ways i) expert elicitation, ii) applying learning algorithms on historical observed values, \cite{chen2012good} or iii) assigning an initial CPT and having it be updated on new observed data\cite{heckerman1995learning}. 

In presence of availability of a fully labeled dataset, method ii) is more appropriate as CPT of the nodes can be learnt from historical values. 
One of the popular algorithms to learn parameters of BNs from historical dataset is the Maximum Likelihood Estimation (MLE) algorithm.
In absence of a large dataset, domain knowledge in form prior beliefs can be incorporated into the learning process.
This is done through the Maximum A Posteriori (MAP) algorithm.
\paragraph{BN for DT}
A possible formulation of a BN for the causal relationships in scenario 1 described in section \ref{sec: motivation} is shown in Figure \ref{fig: scenario1}. 
For our purposes, we propose to construct BNs in three `tiers': i) observed anomaly node (e.g., accelerated failure), 
ii) primary possible explanations of observed failure. 
The nodes denoting parametric trojan inserted, recycled chip, defective chip- all with common child nodes are this type of nodes, 
iii) The set of data items or features that have a correlation with the second tier of nodes. 
As an example, gate dimension, doping recipe, etching depth have correlation with parametric trojans. 
Thus they form the third tier in Figure \ref{fig: scenario1}.
Due to space constraints, we do not show the entire set of features as discussed in section \ref{subsection: security}.
These nodes do not have any parent nodes.

BNs are suited for capturing relationships inherent in backward trust assurance problem for the DT since we would have evidence of some type of anomaly which can be explained by many underlying causes. 
These causal relationships are easily represented by a BN. 
Furthermore, the posterior probabilities $P(T|F), P(R|F)$ and $P(O|F)$ obtained through inference for the BN shown in Figure \ref{fig: scenario1} would assign a probable cause to the query ``Why are chips experiencing accelerated aging in the field?"
Furthermore, if the probabilities such as $P(G|F), P(E|F)$ can be calculated, we would know, within the bounds of a confidence level, that a certain lifecycle stage is where the problem originated from.
For the illustrative BN shown in Figure \ref{fig: scenario1}, the CPT of each of the nodes would be best learnt by encoding the prior belief that recycling is much more likely to be the problem. 
The probabilities each of the node can then be updated by having new data points. 

\subsubsection{Hidden Markov Models}
A HMM is an augmented Markov Chain that models the situation where a sequence of hidden processes influences the outcome of a set of observable events. 
Like Markov Chains, HMMs postulate that the future state of the system can be predicted by knowledge of only the present state of the system. 
Formally, an HMM is characterized by the following set of parameters:
\begin{itemize}
    \item A set of N states $Q=q_1,q_2,...,q_N$
    \item A transition probability matrix $A=a_{11},...,a_{ij},...,a_{NN}$ with each $a_{ij}$ representing the probability of the system transitioning from a state i to state j.
    \item A sequence of observations $O=o_1,o_2, ..., o_T$
    \item The emission probability matrix $B$ which is the probability of a state i generating the observation $o_T$
    \item An initial probability distribution of states $\pi = \pi_1, \pi_2, ...., \pi_N$
\end{itemize}
\paragraph{Fundamental problems of HMMs}
According to Rabiner\cite{rabiner1989tutorial,rabiner1986introduction} there are three fundamental problems that can be answered through HMM modeling: 
\begin{itemize}
    \item[i.] \textbf{Likelihood:} Given a HMM $(A,B)$, the likelihood problem is to determine how likely an observation sequence is to be obtained from that HMM.
    \item[ii.] \textbf{Decoding:} Given an HMM and an obtained observation sequence, the decoding problem is to find the best sequence of hidden states.
    \item[iii.] \textbf{Learning:} The learning problem is to calculate the parameter matrices A and B when presented with an observation sequence.
\end{itemize}

\paragraph{HMM for DT} 

\begin{figure}[!t]
\centering
\subfloat[HMM modeling of `ideal' situation]{\includegraphics[width=0.48\textwidth]{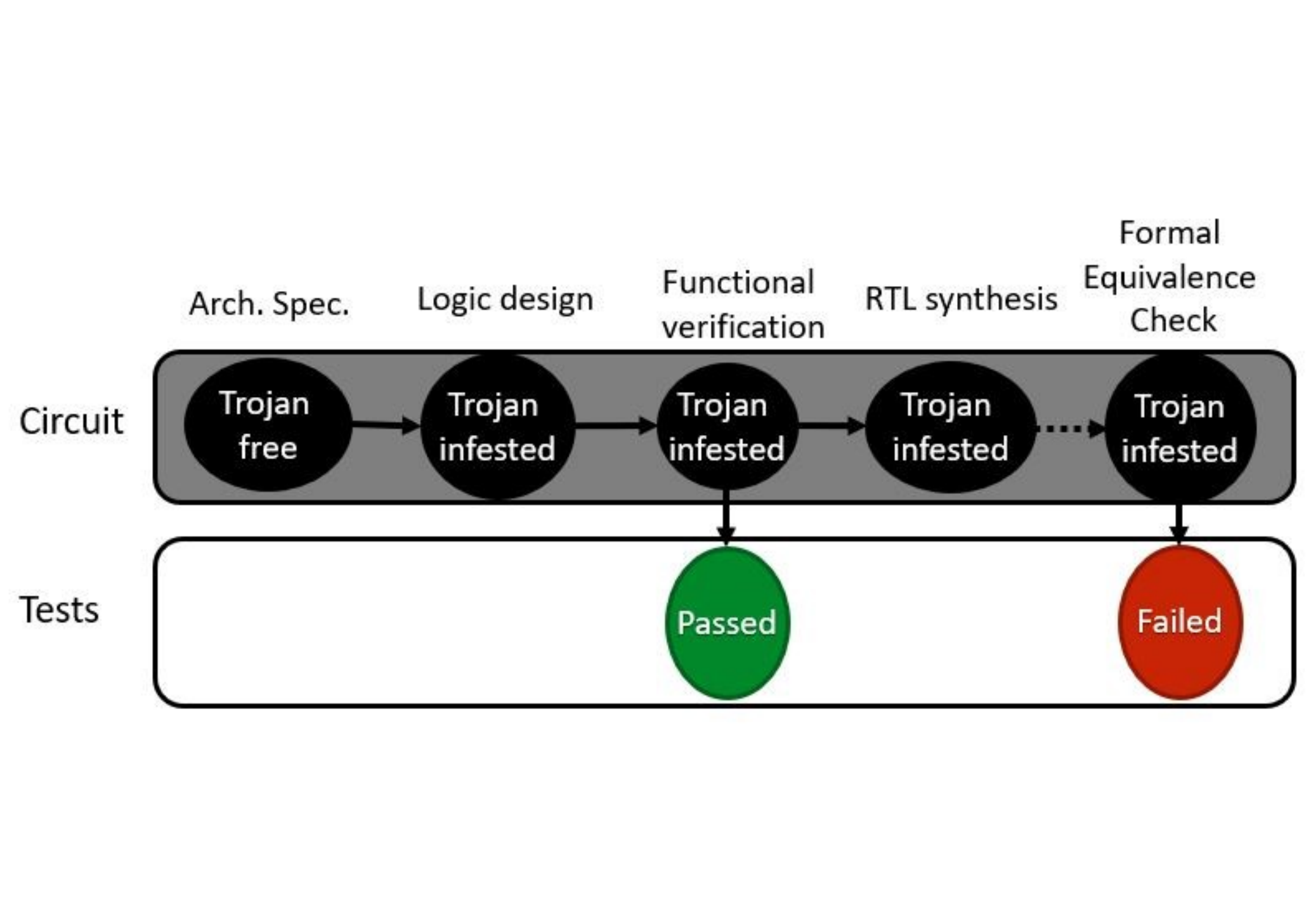}%
\label{fig: ideal}}
\hfil
\subfloat[HMM modeling situation described in scenario 2]{\includegraphics[width=0.48\textwidth]{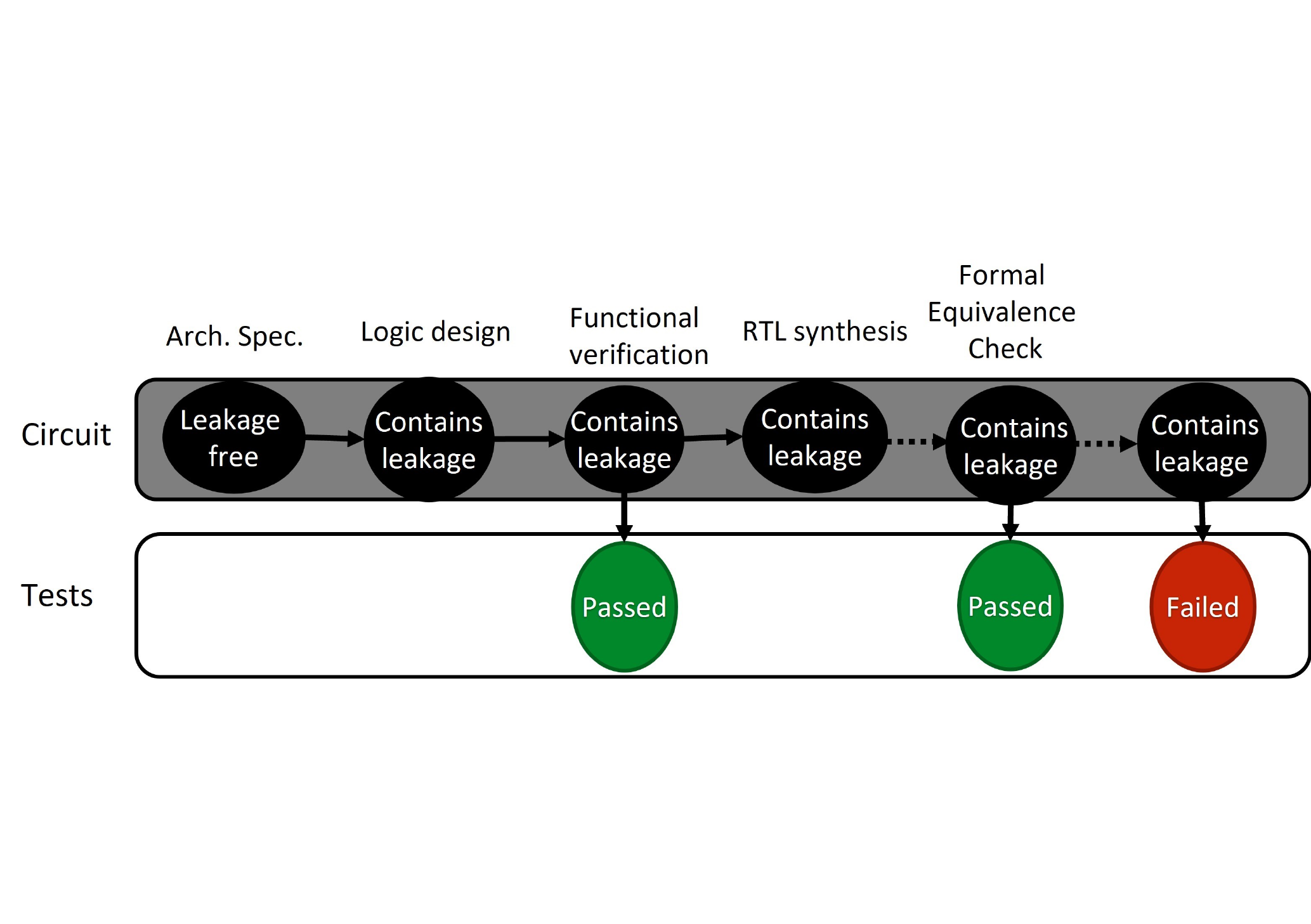}%
\label{fig: actual}}
\caption{HMM modeling for ensuring backward trust in semiconductor lifecycle}
\label{fig: HMM}
\end{figure}

Let us consider scenario 2 and assume that a hardware trojan was inserted.
In an ideal case, one of the pre-silicon verification tests would be able to detect that a trojan was inserted. 
This is shown in Figure \ref{fig: ideal} where we show how the lifecycle process in a case where formal verification was able to detect a trojan insertion, may be modeled as an HMM. 
The actual trojan was inserted in the logic design phase through a 3P soft IP.
This caused the circuit to transition from a trojan free state to trojan infested state.
In this particular threat scenario, the transition of the circuit from trojan free to trojan infested also represents it transitioning from an information leakage free circuit to a leaky circuit.
The argument for modeling hardware security threat scenarios as HMM is that trojan insertion (or a circuit becoming information leakage prone) is a clandestine event and the state of circuit is not observable to us unless a sequence of tests is performed.
The progression of the lifecycle stages constitute the hidden Markov process whereas the sequence of tests performed constitute the observable process which is influenced by the hidden states.
Figure \ref{fig: actual} shows the more likely scenario.
Here, we model the fact that none of the presilicon verification tests were successful (indicated by a `passed' observation).
The anomalous behavior i.e., the leakage of the key was detected at deployment stage through JTAG test.

\begin{figure}[!t]
\centering
\subfloat[HMM learning for scenario 2]{\includegraphics[width=0.35\textwidth]{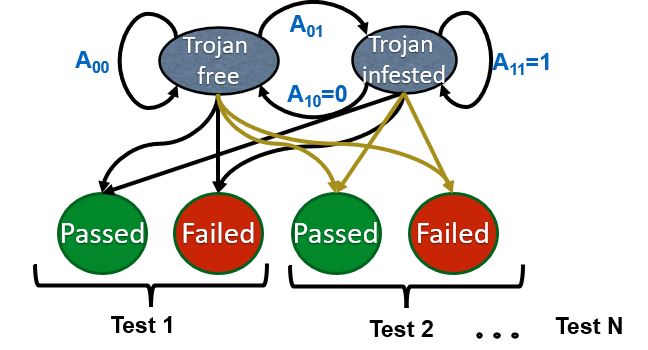}%
\label{fig: hmm_learn}}
\hfil
\subfloat[HMM learning and inference process]{\includegraphics[width=0.4\textwidth]{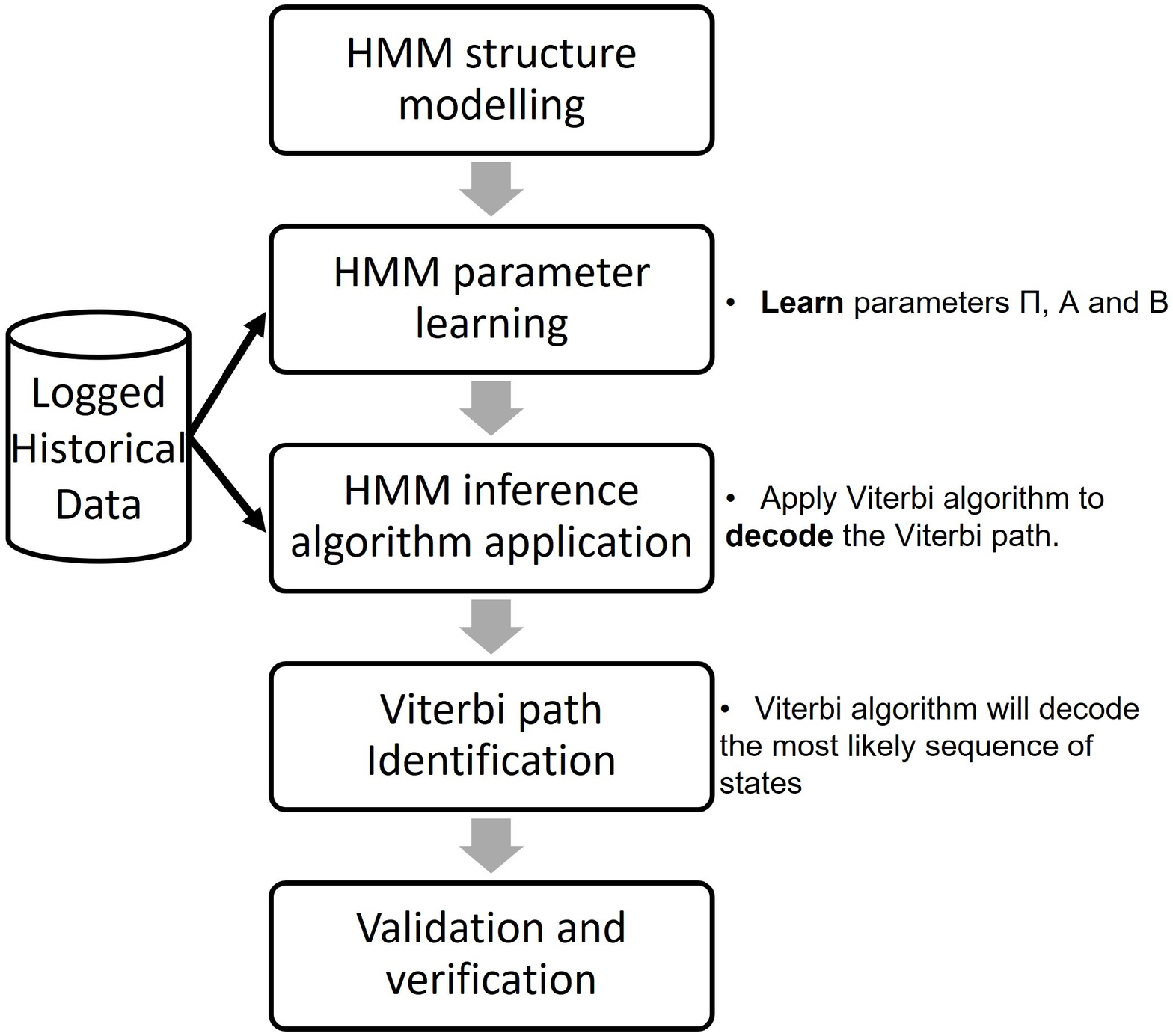}%
\label{fig: hmm_process}}
\caption{HMM learning and decoding for ensuring backward trust}
\label{fig: HMM2}
\end{figure}

For security assurance through the proposed DT, the root cause problem we are interested in solving can be framed as the Decoding problem of HMM.
In Figure \ref{fig: HMM}, the Viterbi algorithm \cite{forney1973viterbi} may be used to find the most probable sequence of events. 
If the modeling is successful, we can reasonably expect the constructed HMM model to infer that the circuit changed state in the logic design phase.
From the description of the Decoding problem given earlier, to apply Viterbi algorithm prior knowledge of the transition and emission matrices is needed. 
Obviously, it is challenging to know what is the emission probability from a trojan free state to a functional test giving a `pass' result.
For this reason, intially from historical observation data analysis, the matrices A and B must be learned. 
Here, certain facts should be noted. 
The probability that a trojan infested circuit would transition to a trojan free circuit is 0. 
Therefore, certain transition probabilities can be assigned from domain knowledge whereas others would need to be learned.
The learning algorithm will also be needed to be adapted for such that only the parameters that are not set by domain knowledge is learnt.
The entire process flow required for HMM realization is shown in Figure \ref{fig: hmm_process}. 
The logged historical data acquired from previous progressions of a device through lifecycle steps can be used to learn the transition and emission probabilities. 
\subsubsection{Markov Logic Networks}
\label{MLN}
An MLN is a SRL model that attributes a weight to set of first-order logic formulae known as the knowledge base (KB) \cite{richardson2006markov}. 
For any domain, a KB written in first order formulae can succinctly represent prior domain knowledge. 
Constants, variables, functions, and predicates are four type of symbols that are used to build formulas in first-order logic. 
Constant symbols are used to represent items within a certain scope of interest. 
Variable symbols span the domain's objects. 
Function symbols denote mappings between tuples of objects and individual objects. 
Predicate symbols describe relationships between items or properties of objects in the domain. 
An interpretation describes which symbols are used to represent which objects, functions, and relations in the domain. 
Certain logical connectives and quantifiers (e.g., $\wedge,\vee,\implies,\exists$) are used to construct and qualify these formulae.

However, first-order logic by itself is not suitable to deal with uncertainty inherent in semiconductor lifecycle.
MLNs add the capacity to deal rationally with uncertainty, and to tolerate unsure and conflicting knowledge, by constructing a Markov network utilizing each grounding of a formula in the KB. 
More specifically, MLN is an instantiation of a Markov Random Field (MRF) where groundings of the predicate (atom) constitute the nodes and groundings of the formulae (clauses) represent the features.
The atoms in the field are then described by the joint probability distribution:
\begin{equation}
 P(X=x)=\frac{1}{Z}exp(\sum_i w_i n_i(x))    
\end{equation}
where $w_i$ is the weight associated with formula ith formula $f_i$, Z is a constant, and $n_i(x)$ is the number of true groundings of the formula $f_i$ in the world $x$. 
A world corresponds to a small subset of all possible groundings of atoms.

The formulae in the KB are usually designed by experts in the field \cite{lowd2007efficient}.
The weight associated with each formula brings about the notion of `hard' and `soft' formulae.
A hard formula has a weight of $+\infty$ or $-\infty$ signifying always true and false respectively.
A soft formula on the other hand may have weights in between signifying the `degree' of truth in them.
This incorporation of weights is what would allow us to deal with uncertainty inherent in root cause analysis problem of ensuring backward trust.

Additionally, two types of statistical inference is possible in MLNs. 
Given some evidence, Maximum a posteriori inference (MAP) finds most probable world. 
The second type of inference, namely marginal inference, calculates the posterior probability distribution over the values of all variables. 
We are interested in MAP since in backward trust the DT would be presented with some evidence of anomaly and it would have to find the most probable world in which the atom are satisfied.

\subsection{Anomaly Detection}
\label{subsec: Anomaly}
The DT has another important component to its digital analytics engine and that is the anomaly detection methods collectively referred to as \emph{Data analytics algorithm} in Figure \ref{fig: DT_full}.
In scenario I, the initial observed anomaly is a report on accelerated failure of the chip which may be available from customer feedback.
However, in the more general case, the suspected behavior may not be so easy to confirm.
For example, authors in \cite{skorobogatov2012breakthrough} had to devise their own Pipeline Emisson Analytics to discover that a military grade commercially available FPGA had a possible backdoor in it.
Depending on the suspected anomalous behavior, the analysis technique would have to be different.
As a result, the DT requires a preliminary data analytics algorithm that would find the anomaly in the suspected data stream and subsequently infer the list of possible causes.
These data analytics algorithm can be something as simple as visual inspection of a high resolution microscopy image to more involved analytics like time series analysis methods, statistical signal processing algorithms or machine learning and deep learning models.

Furthermore, it is the evidence of anomaly gathered from various different algorithms that drives the training database of the SRL model.
The data collected from various stages of the lifecycle are extremely varied in nature. 
Data can be textual such as the RTL code from design phase, or the Jobdeck file from mask writing phase.
The fabrication floor data such as gate dimension, etching depth, doping dosage are numerical data.
As the type of data is different, the analytics algorithm would also be different.
This is also the reason why we have included a `feature extractor' block that takes these disparate types of data and translates them into legible values understandable by core reasoning algorithms.

We discuss some of the data analytics that can be performed to find traces of anomaly in different types of data.
Let us consider threat scenario I. 
To find evidence of anomaly in data such as doping dosage or etching depth, a method such as the one proposed in\cite{laptev2015generic} may be used.
If an insider, for example, changes the recipe of the doping dosage this would be detectable by the Kernel Density Estimation (KDE) and the Kullback-Liebler divergence based change point anomaly detection.
A mismatch between the OCM specification and the DUT's observed measurements represent anomaly in data like lead, ball or pin dimensions, chemical composition, straightness, alignment etc. 
To identify it, a simple visual inspection from a X-ray or SEM image may suffice.
For more involved techniques proposed in literature may also be used for achieving greater confidence in the obtained evidence.
The 4D enhanced SEM imaging proposed in \cite{shahbazmohamadi2014advanced}, the machine vision and advanced image processing methods demonstrated in \cite{baygin2017machine,ghosh2018recycled} represent such more sophisticated methods.
For data such as wafer and master results record, a cross match of the STDF file with expected values would reveal anomalies. 
This cross match may be automated through natural language processing algorithms.
For scenario 2, the data to inspect would be the RTL and netlist files in the design phase in addition to the fabrication stage data discussed for scenario 1.

\begin{figure}[!t]
\centering
\subfloat[Plot for trojan free circuit]{\includegraphics[width=0.45\textwidth]{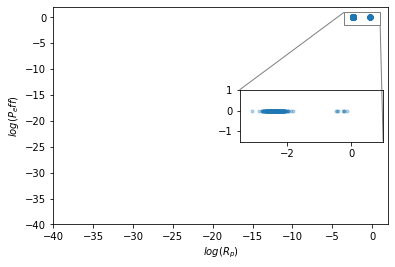}%
\label{fig: trojanfree}}
\hfil
\subfloat[Plot for trojan infested circuit]{\includegraphics[width=0.45\textwidth]{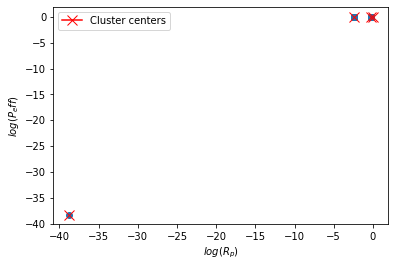}%
\label{fig: trojaninfested}}
\caption{Anomaly can be detected in a plot of relative branching probability vs effective branching probability when comparing trojan infested and trojan free versions of AES-T1300. The cluster centers were determined by Mean Shift algorithm.}
\label{fig: anomaly_troj}
\end{figure}

In case a information leaking hardware trojan was inserted, a plot of relative branching probability ($R_p$) vs effective branching probability ($P_{eff}$), as defined in \cite{choo2020register}, may demonstrate an anomaly. 
For example in Figure \ref{fig: anomaly_troj}, we show the plot of $R_p$ vs $P_{eff}$ for the trojan-free version of the circuit AES-T1300 (an example of an information leaking trojan available from the TrustHub database\cite{shakya2017benchmarking,salmani2013design}) as compared to the trojan infested version. 
There is a very distant cluster of branches corresponding to the trojan trigger branches as measured by cluster centers derived from the Mean Shift algorithm.
The structural features of the circuit were extracted by a RTL parser which is fulfilling the role of `Feature extractor or normalizer' in Figure \ref{fig: DT_full}.
We have discussed previously in section \ref{sec: HAV} how traditional methods for testing for unintentional information leakage and information leaking hardware trojan provide low coverage and low confidence. 
New testing methods such as concolic testing \cite{meng2021rtl}, RTL level IFT\cite{ardeshiricham2017register} have been proposed that offer a greater deal of certainty in finding implicit and explicit information flows in the design.
Any one of these testing methods may be used to find evidences of information leakage in the design which would serve as evidence of anomaly.
 
We answer this question in \ref{subsec: forward} through the explanation of the second proposed functionality of the DT framework, i.e., forward trust analysis.

\subsection{Forward Trust Analysis}
\label{subsec: forward}
We have discussed how existing solutions in hardware security do not scale very well either because they address only one or two specific vulnerability or because they require the embedding of overhead incurring sensors and/or modification of existing process flows.
Enabling forward trust refers to the scalability and adaptability of the DT framework. 
Scalability means that the proposed framework would be usable to address all possible attack vectors instead of a subset of the attack vectors.
Adaptability refers to the ability of DT to address emerging threats by effectively carrying out root cause analysis even unforeseen emerging threats.
As ease of accessibility to computation resources keep increasing and the semiconductor industry becomes more horizontally globalized, new threats would continue to emerge.
Adaptability addresses this issue by leveraging the fact that as hardware security threats evolve over time, so do our collective understanding of their underlying causes and the corresponding anomaly detection algorithms.

Scalability is possible in the proposed DT framework by simply translating the domain knowledge to an equivalent KB for MLN or to a DAG for characterizing the BN and HMM.
The same basic principals would apply as illustrated through the running scenarios presented in paper for other attack vectors such as side channel and fault injection attacks.
Adaptability is also an inherent feature of the SRL models discussed previously. 
Let us consider an example. 
Researchers in \cite{kison2019security} demonstrated a novel trojan that circumvent existing verification efforts and introduces capacitive crosstalk in between metal layers to realize its payload.
Now, to enable root cause analysis of this scenario all we are required to do is to add a parent node to the child node `parametric trojan inserted' called `rerouting the metal layers'. 
Similarly for MLNs, this would amount to a new grounding of the formulae in the KB. 
Without changing the core algorithms drastically, we can still model new threats so long as the underlying causes are known.
Furthermore, as the individual anomaly detection algorithms are `detached' from the core functionality of the root cause analysis AI model, they can be updated as better anomaly detection algorithms emerge.
One of the examples in context of information flow tracking was already discussed in \ref{subsec: Anomaly}.
\subsection{Process Flow for Trust and Security Analysis}
\label{subsection: Process Flow}
\begin{figure}[!t]
\centering
\includegraphics[width=3.5in]{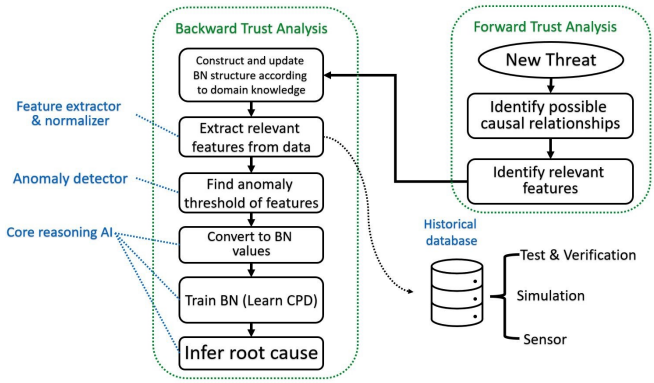}

\caption{Backward and Forward Trust Analysis Process Flows}
\label{fig: process_flow}

\end{figure}

We now summarize the entire process flow with the help of Figure \ref{fig: process_flow} and scenarios mentioned earlier. 
For the sake of simplicity, we constrain our discussion of the process flow with the assumption that BN is the core AI algorithm for root cause analysis.
We have already shown the BN network structure for scenario 1 and 3 (shown in Figure \ref{fig: scenario1}).
Next, we have the knowledge of relevant features of the threat model, which the reader can consult from section\ref{subsection: security}.
For example, we can extract relevant features from RTL and GLN, an RTL and netlist parser or using commercial EDA tools in scenario 2.
The anomaly detector would then find the associated threshold of the features that signify an anomaly.
A BN with binary outcomes for each node can be constructed by determining whether corresponding features exceed the threshold.
For each entry in the database, the BN can then be trained to learn the CPT parameters.
This builds the initial CPT of the nodes in the BN.
Next, the features of design under test can be extracted in the same process and CPT parameters can be updated according to a learning rate. 
Lastly, inference from BN can point us to the probable cause.
The process thus far described constitutes the Backward Trust Analysis.
For adaptability to new threats, the domain knowledge needs to be updated. 
This amounts to updating the BN structure by adding the proper child or parent nodes and identifying relevant features.
We have already discussed one example of this earlier in \ref{subsec: forward}.
Once this has been updated, we can proceed as before in Backward Trust Analysis to infer the origin of the threat.

\section{Challenges and Future Research Avenues}
\label{sec: challenges}
We have laid a clear outline of the algorithms and components required to realize a digital twin for end-to-end semiconductor lifecycle management in the preceding sections. 
A full implementation of the proposed framework when working on real datasets will present several challenges pertaining to optimization, computational complexity, logistics of data transfer and adopted technologies. 
We discuss these challenges in this section and provide some indications on how to overcome them.
\subsection{Defining the Virtual Environment}
\label{subsec: virtual}
According to the definition of digital twins used for lifecycle management, there are five principal dimensions: physical part, digital part, data, service and connection\cite{tao2018digital}. 
For secure semiconductor lifecycle the physical part is a process instead of a physical object. 
More specifically, we are modeling the lifecycle of a physical object as it progresses through different stages.
We have also defined the data that the DT will leverage, the services it will provide as well namely, forward and backward trust analysis. 
In DT literature, virtual environments are also defined as the containers of the digitized twin. 
It is most often a cloud platform or data warehouse to host the relevant database and run AI models on said database although it need not be\cite{jones2020characterising}.
In the preceding, we have not defined the cloud platform to host the database and the digitized twin. 
Platforms such AWS or Microsoft Azure can be used for this purpose.
\subsection{Data Acquisition and Security}
\label{subsec: data security}
For the first scenario mentioned in section \ref{sec: motivation}, we illustrated which data can be used to assign a probable provenance to a hardware attack in section \ref{subsection: security}.
However, this ignores the fact that the foundry may not be willing to share its manufacturing data.
This does not automatically diminish the usefulness of our proposed framework.
In such a case, destructive reverse engineering of a population of the used available chips will be required to find evidences of CD change in devices which indicates an parametric trojan\cite{vashistha2018detecting}.
Similarly, other innovative data mining and acquisition techniques will need to be employed in case certain data items are not available to the entity with the access to the DT.

Furthermore, secure acquisition of relevant data items is an avenue we have not discussed. 
In creating a cyber physical system, there is always the risk of false data injection attacks which may be carried out by an insider through a network.
Falsification of test data at the fabrication step, for example, may be traced through the Audit Trail Record of STDF database. 
At other steps, or through the network, tracing data injection attacks would require a completely separate approach which is not explored in this paper.

\subsection{State Space Complexity}
\label{subsec: state space}
The Bayesian Network presented in section \ref{subsection: srl} shows how a child may have three parent nodes. 
According to\cite{marcot2006guidelines}, this presents a significant challenge to keep the task of populating the associated CPTs tractable. 
Indeed, for many threat vectors, each child node may have quite a few more than three parent nodes.
A similar performance speed optimization problem may be encountered for modeling cybersecurity threats through MLN KB. 
As the domain of discourse for predicates may become unacceptably large for quick convergence and training, parallelization\cite{beedkar2013fully} of the inference process or preprocessing of the data\cite{shavlik2009speeding} will be required in more unrestricted threat scenarios.

\subsection{Model Optimization}
\label{subsec: model}
Due to the complexity of the model and potentially multiple faceted causal relationships, it might become computationally expensive and time consuming for a SRL model to converge to a solution.
Additionally, since data driven secure lifecycle management approaches in the semiconductor industry is still in its infancy, there might be limited availability of labeled datasets. 
Therefore, a multi-factor optimization of the network would be needed considering the large state space, learning parameter constraints, and limitations of the enabling technologies applied.
Design space exploration and hyperparameter tuning thus becomes a very significant challenge for more involved and complex threat models.

\subsection{Model Upgradeability}
\label{subsec: upgradeability}
Forward trust analysis is an indispensable cornerstone of the proposed DT functionality.
For many novel threats reported in literature, the underlying causal relationships are known to an extent as soon as they are discovered.
For example, in the first report of Spectre attacks authors clearly identified that they were exploiting speculative execution vulnerability in the hardware implementation of the processors.
However, in case of many zero day threats which are not discovered by researchers beforehand, the causal relationships may not be readily understood.
In such a case, the core SRL model structure would need to learnt from the data available.
One example of how it can be done is the learning of Bayesian network structure through evolutionary algorithms\cite{daly2009learning}.

\subsection{Threat Mitigation}
\label{subsec: mitigation}
Although the proposed framework offers scalability to future threats through inherently expandable and updatable AI algorithms, it does not provide any provision for mitigation of threats by itself. 
Once the root cause has been identified, the appropriate countermeasure to be taken is left to the discretion of the defender.
Countermeasures can be taken across both hardware and software layers of the computing stack.
Implementing hardware countermeasures once it has been deployed is extremely difficult especially for ASIC platform.
For FPGA platform, the uploaded design bitstream may be comparatively easily updated with proper measures although monetary cost of taking a system offline to do so is a great concern. 
Hardware patching\cite{nath2018system} has been proposed recently although such efforts remain in infancy both in terms of adoption and usability.
Embedded FPGA\cite{beyrouthy2007novel,saha2020fpga} and programmable hardware monitors\cite{delshadtehrani2020phmon} represent other alternatives which require further investigation.

Software patching remains the go to option for addressing hardware security concerns due to the relative ease of their implementation.
For instance, even though both Meltdown and Spectre attacks exploit vulnerabilities present in hardware, vast majority of defenses deployed infield have been software countermeasures. 
Effective implementation of threat mitigation of a wide range of hardware security concerns may have to be tackled through both hardware-software `co-upgrading' similar to hardware-software co-design and co-verification practices being proposed in literature.
We plan to visit these questions in our future works on the proposed DT framework.
\section{Conclusion}
\label{sec: conc}
We proposed a novel approach for end-to-end secure semiconductor lifecycle and supply chain management.
We discussed how this approach is founded on taking advantage of causal relationships between available data from semiconductor lifecycle flows and hardware attack vectors.
We outlined all the necessary components for realizing and implementing a digital twin that leverages these relationships.
The relationships between three hardware attack vectors and associated data items were highlighted.
The steps required to reconcile disparate types of data items with core reasoning SRL models were also delineated.
We also discussed some of the significant challenges in realizing this framework and potential methods that can be adopted to address some of them.
We hope this paper stimulates further research in end-to-end data driven lifecycle management of semiconductor devices.

\bibliographystyle{IEEEtran}
\bibliography{IEEEabrv,mybib.bib,lfi.bib}

\begin{thebibliography}{100}
\providecommand{\url}[1]{#1}
\csname url@samestyle\endcsname
\providecommand{\newblock}{\relax}
\providecommand{\bibinfo}[2]{#2}
\providecommand{\BIBentrySTDinterwordspacing}{\spaceskip=0pt\relax}
\providecommand{\BIBentryALTinterwordstretchfactor}{4}
\providecommand{\BIBentryALTinterwordspacing}{\spaceskip=\fontdimen2\font plus
\BIBentryALTinterwordstretchfactor\fontdimen3\font minus
  \fontdimen4\font\relax}
\providecommand{\BIBforeignlanguage}[2]{{%
\expandafter\ifx\csname l@#1\endcsname\relax
\typeout{** WARNING: IEEEtran.bst: No hyphenation pattern has been}%
\typeout{** loaded for the language `#1'. Using the pattern for}%
\typeout{** the default language instead.}%
\else
\language=\csname l@#1\endcsname
\fi
#2}}
\providecommand{\BIBdecl}{\relax}
\BIBdecl

\bibitem{khattri2012hsdl}
H.~Khattri, N.~K.~V. Mangipudi, and S.~Mandujano, ``Hsdl: A security
  development lifecycle for hardware technologies,'' in \emph{2012 IEEE
  International Symposium on Hardware-Oriented Security and Trust}.\hskip 1em
  plus 0.5em minus 0.4em\relax IEEE, 2012, pp. 116--121.

\bibitem{dessouky2019hardfails}
G.~Dessouky, D.~Gens, P.~Haney, G.~Persyn, A.~Kanuparthi, H.~Khattri, J.~M.
  Fung, A.-R. Sadeghi, and J.~Rajendran, ``Hardfails: Insights into
  software-exploitable hardware bugs,'' in \emph{28th USENIX Security
  Symposium}, 2019, pp. 213--230.

\bibitem{sami2021end}
M.~S. U.~I. Sami, F.~Rahman, F.~Farahmandi, A.~Cron, M.~Borza, and
  M.~Tehranipoor, ``End-to-end secure soc lifecycle management,'' in \emph{2021
  58th ACM/IEEE Design Automation Conference (DAC)}.\hskip 1em plus 0.5em minus
  0.4em\relax IEEE, 2021, pp. 1295--1298.

\bibitem{ray2017system}
S.~Ray, E.~Peeters, M.~M. Tehranipoor, and S.~Bhunia, ``System-on-chip platform
  security assurance: Architecture and validation,'' \emph{Proceedings of the
  IEEE}, vol. 106, no.~1, pp. 21--37, 2017.

\bibitem{xu2019electronics}
X.~Xu, F.~Rahman, B.~Shakya, A.~Vassilev, D.~Forte, and M.~Tehranipoor,
  ``Electronics supply chain integrity enabled by blockchain,'' \emph{ACM
  Transactions on Design Automation of Electronic Systems (TODAES)}, vol.~24,
  no.~3, pp. 1--25, 2019.

\bibitem{zhang2018chip}
D.~Zhang, X.~Wang, M.~T. Rahman, and M.~Tehranipoor, ``An on-chip dynamically
  obfuscated wrapper for protecting supply chain against ip and ic piracies,''
  \emph{IEEE Transactions on Very Large Scale Integration (VLSI) Systems},
  vol.~26, no.~11, pp. 2456--2469, 2018.

\bibitem{jones2020characterising}
D.~Jones, C.~Snider, A.~Nassehi, J.~Yon, and B.~Hicks, ``Characterising the
  digital twin: A systematic literature review,'' \emph{CIRP Journal of
  Manufacturing Science and Technology}, vol.~29, pp. 36--52, 2020.

\bibitem{glaessgen2012digital}
E.~Glaessgen and D.~Stargel, ``The digital twin paradigm for future nasa and us
  air force vehicles,'' in \emph{53rd AIAA/ASME/ASCE/AHS/ASC structures,
  structural dynamics and materials conference 20th AIAA/ASME/AHS adaptive
  structures conference 14th AIAA}, 2012, p. 1818.

\bibitem{fuller2020digital}
A.~Fuller, Z.~Fan, C.~Day, and C.~Barlow, ``Digital twin: Enabling
  technologies, challenges and open research,'' \emph{IEEE access}, vol.~8, pp.
  108\,952--108\,971, 2020.

\bibitem{bao2019modelling}
J.~Bao, D.~Guo, J.~Li, and J.~Zhang, ``The modelling and operations for the
  digital twin in the context of manufacturing,'' \emph{Enterprise Information
  Systems}, vol.~13, no.~4, pp. 534--556, 2019.

\bibitem{madni2019leveraging}
A.~M. Madni, C.~C. Madni, and S.~D. Lucero, ``Leveraging digital twin
  technology in model-based systems engineering,'' \emph{Systems}, vol.~7,
  no.~1, p.~7, 2019.

\bibitem{bachelor2019model}
G.~Bachelor, E.~Brusa, D.~Ferretto, and A.~Mitschke, ``Model-based design of
  complex aeronautical systems through digital twin and thread concepts,''
  \emph{IEEE Systems Journal}, vol.~14, no.~2, pp. 1568--1579, 2019.

\bibitem{martinez2018automatic}
G.~S. Mart{\'\i}nez, S.~Sierla, T.~Karhela, and V.~Vyatkin, ``Automatic
  generation of a simulation-based digital twin of an industrial process
  plant,'' in \emph{IECON 2018-44th Annual Conference of the IEEE Industrial
  Electronics Society}.\hskip 1em plus 0.5em minus 0.4em\relax IEEE, 2018, pp.
  3084--3089.

\bibitem{kritzler2019digital}
M.~Kritzler, J.~Hodges, D.~Yu, K.~Garc{\'\i}a, H.~Shukla, and F.~Michahelles,
  ``Digital companion for industry,'' in \emph{Companion Proceedings of The
  2019 World Wide Web Conference}, 2019, pp. 663--667.

\bibitem{qi2018digital}
Q.~Qi and F.~Tao, ``Digital twin and big data towards smart manufacturing and
  industry 4.0: 360 degree comparison,'' \emph{Ieee Access}, vol.~6, pp.
  3585--3593, 2018.

\bibitem{liu2021review}
M.~Liu, S.~Fang, H.~Dong, and C.~Xu, ``Review of digital twin about concepts,
  technologies, and industrial applications,'' \emph{Journal of Manufacturing
  Systems}, vol.~58, pp. 346--361, 2021.

\bibitem{wurm2016introduction}
J.~Wurm, Y.~Jin, Y.~Liu, S.~Hu, K.~Heffner, F.~Rahman, and M.~Tehranipoor,
  ``Introduction to cyber-physical system security: A cross-layer
  perspective,'' \emph{IEEE Transactions on Multi-Scale Computing Systems},
  vol.~3, no.~3, pp. 215--227, 2016.

\bibitem{guin2016fortis}
U.~Guin, Q.~Shi, D.~Forte, and M.~M. Tehranipoor, ``Fortis: a comprehensive
  solution for establishing forward trust for protecting ips and ics,''
  \emph{ACM transactions on design automation of electronic systems (TODAES)},
  vol.~21, no.~4, pp. 1--20, 2016.

\bibitem{dabrowski2014towards}
A.~Dabrowski, H.~Hobel, J.~Ullrich, K.~Krombholz, and E.~Weippl, ``Towards a
  hardware trojan detection cycle,'' in \emph{2014 Ninth international
  conference on availability, Reliability and Security}.\hskip 1em plus 0.5em
  minus 0.4em\relax IEEE, 2014, pp. 287--294.

\bibitem{tehranipoor2014integrated}
M.~Tehranipoor, H.~Salmani, and X.~Zhang, ``Integrated circuit
  authentication,'' \emph{Switzerland: Springer, Cham. doi}, vol.~10, pp.
  978--3, 2014.

\bibitem{bhunia2013protection}
S.~Bhunia, M.~Abramovici, D.~Agrawal, P.~Bradley, M.~S. Hsiao, J.~Plusquellic,
  and M.~Tehranipoor, ``Protection against hardware trojan attacks: Towards a
  comprehensive solution,'' \emph{IEEE Design \& Test}, vol.~30, no.~3, pp.
  6--17, 2013.

\bibitem{bhunia2018hardware}
S.~Bhunia and M.~Tehranipoor, \emph{Hardware security: a hands-on learning
  approach}.\hskip 1em plus 0.5em minus 0.4em\relax Morgan Kaufmann, 2018.

\bibitem{xiao2016hardware}
K.~Xiao, D.~Forte, Y.~Jin, R.~Karri, S.~Bhunia, and M.~Tehranipoor, ``Hardware
  trojans: Lessons learned after one decade of research,'' \emph{ACM
  Transactions on Design Automation of Electronic Systems (TODAES)}, vol.~22,
  no.~1, pp. 1--23, 2016.

\bibitem{guin2014low}
U.~Guin, X.~Zhang, D.~Forte, and M.~Tehranipoor, ``Low-cost on-chip structures
  for combating die and ic recycling,'' in \emph{2014 51st ACM/EDAC/IEEE Design
  Automation Conference (DAC)}.\hskip 1em plus 0.5em minus 0.4em\relax IEEE,
  2014, pp. 1--6.

\bibitem{guin2014counterfeit}
U.~Guin, K.~Huang, D.~DiMase, J.~M. Carulli, M.~Tehranipoor, and Y.~Makris,
  ``Counterfeit integrated circuits: A rising threat in the global
  semiconductor supply chain,'' \emph{Proceedings of the IEEE}, vol. 102,
  no.~8, pp. 1207--1228, 2014.

\bibitem{wang2012representative}
S.~Wang, J.~Chen, and M.~Tehranipoor, ``Representative critical reliability
  paths for low-cost and accurate on-chip aging evaluation,'' in
  \emph{Proceedings of the International Conference on Computer-Aided Design},
  2012, pp. 736--741.

\bibitem{guin2014comprehensive}
U.~Guin, D.~DiMase, and M.~Tehranipoor, ``A comprehensive framework for
  counterfeit defect coverage analysis and detection assessment,''
  \emph{Journal of Electronic Testing}, vol.~30, no.~1, pp. 25--40, 2014.

\bibitem{alkabani2007active}
Y.~Alkabani and F.~Koushanfar, ``Active hardware metering for intellectual
  property protection and security.'' in \emph{USENIX security symposium},
  2007, pp. 291--306.

\bibitem{huang2015recycled}
K.~Huang, Y.~Liu, N.~Korolija, J.~M. Carulli, and Y.~Makris, ``Recycled ic
  detection based on statistical methods,'' \emph{IEEE transactions on
  computer-aided design of integrated circuits and systems}, vol.~34, no.~6,
  pp. 947--960, 2015.

\bibitem{rajendran2016formal}
J.~Rajendran, A.~M. Dhandayuthapany, V.~Vedula, and R.~Karri, ``Formal security
  verification of third party intellectual property cores for information
  leakage,'' in \emph{2016 29th International conference on VLSI design and
  2016 15th international conference on embedded systems (VLSID)}.\hskip 1em
  plus 0.5em minus 0.4em\relax IEEE, 2016, pp. 547--552.

\bibitem{guo2019qif}
X.~Guo, R.~G. Dutta, J.~He, M.~M. Tehranipoor, and Y.~Jin, ``Qif-verilog:
  Quantitative information-flow based hardware description languages for
  pre-silicon security assessment,'' in \emph{2019 IEEE International Symposium
  on Hardware Oriented Security and Trust (HOST)}.\hskip 1em plus 0.5em minus
  0.4em\relax IEEE, 2019, pp. 91--100.

\bibitem{wang2021sofi}
H.~Wang, H.~Li, F.~Rahman, M.~M. Tehranipoor, and F.~Farahmandi, ``Sofi:
  Security property-driven vulnerability assessments of ics against
  fault-injection attacks,'' \emph{IEEE Transactions on Computer-Aided Design
  of Integrated Circuits and Systems}, 2021.

\bibitem{nahiyan2016avfsm}
A.~Nahiyan, K.~Xiao, K.~Yang, Y.~Jin, D.~Forte, and M.~Tehranipoor, ``Avfsm: A
  framework for identifying and mitigating vulnerabilities in fsms,'' in
  \emph{2016 53nd ACM/EDAC/IEEE Design Automation Conference (DAC)}.\hskip 1em
  plus 0.5em minus 0.4em\relax IEEE, 2016, pp. 1--6.

\bibitem{nahiyan2018security}
A.~Nahiyan, F.~Farahmandi, P.~Mishra, D.~Forte, and M.~Tehranipoor,
  ``Security-aware fsm design flow for identifying and mitigating
  vulnerabilities to fault attacks,'' \emph{IEEE Transactions on Computer-aided
  design of integrated circuits and systems}, vol.~38, no.~6, pp. 1003--1016,
  2018.

\bibitem{park2019leveraging}
J.~Park, F.~Rahman, A.~Vassilev, D.~Forte, and M.~Tehranipoor, ``Leveraging
  side-channel information for disassembly and security,'' \emph{ACM Journal on
  Emerging Technologies in Computing Systems (JETC)}, vol.~16, no.~1, pp.
  1--21, 2019.

\bibitem{shan2019machine}
W.~Shan, S.~Zhang, J.~Xu, M.~Lu, L.~Shi, and J.~Yang, ``Machine learning
  assisted side-channel-attack countermeasure and its application on a 28-nm
  aes circuit,'' \emph{IEEE Journal of Solid-State Circuits}, vol.~55, no.~3,
  pp. 794--804, 2019.

\bibitem{nahiyan2020script}
A.~Nahiyan, J.~Park, M.~He, Y.~Iskander, F.~Farahmandi, D.~Forte, and
  M.~Tehranipoor, ``Script: a cad framework for power side-channel
  vulnerability assessment using information flow tracking and pattern
  generation,'' \emph{ACM Transactions on Design Automation of Electronic
  Systems (TODAES)}, vol.~25, no.~3, pp. 1--27, 2020.

\bibitem{tehranipoor2015counterfeit}
M.~M. Tehranipoor, U.~Guin, and D.~Forte, ``Counterfeit integrated circuits,''
  in \emph{Counterfeit Integrated Circuits}.\hskip 1em plus 0.5em minus
  0.4em\relax Springer, 2015, pp. 15--36.

\bibitem{shiyanovskii2010process}
Y.~Shiyanovskii, F.~Wolff, A.~Rajendran, C.~Papachristou, D.~Weyer, and
  W.~Clay, ``Process reliability based trojans through nbti and hci effects,''
  in \emph{2010 NASA/ESA Conference on Adaptive Hardware and Systems}.\hskip
  1em plus 0.5em minus 0.4em\relax IEEE, 2010, pp. 215--222.

\bibitem{kashyap2021silicon}
R.~Kashyap, ``Silicon lifecycle management (slm) with in-chip monitoring,'' in
  \emph{2021 IEEE International Reliability Physics Symposium (IRPS)}.\hskip
  1em plus 0.5em minus 0.4em\relax IEEE, 2021, pp. 1--4.

\bibitem{khalid2020simcom}
F.~Khalid, S.~R. Hasan, O.~Hasan, and M.~Shafique, ``Simcom: Statistical
  sniffing of inter-module communications for runtime hardware trojan
  detection,'' \emph{Microprocessors and Microsystems}, vol.~77, p. 103122,
  2020.

\bibitem{lin2009moles}
L.~Lin, W.~Burleson, and C.~Paar, ``Moles: Malicious off-chip leakage enabled
  by side-channels,'' in \emph{2009 IEEE/ACM International Conference on
  Computer-Aided Design-Digest of Technical Papers}.\hskip 1em plus 0.5em minus
  0.4em\relax IEEE, 2009, pp. 117--122.

\bibitem{tao2018digital}
F.~Tao, H.~Zhang, A.~Liu, and A.~Y. Nee, ``Digital twin in industry:
  State-of-the-art,'' \emph{IEEE Transactions on Industrial Informatics},
  vol.~15, no.~4, pp. 2405--2415, 2018.

\bibitem{bitton2018deriving}
R.~Bitton, T.~Gluck, O.~Stan, M.~Inokuchi, Y.~Ohta, Y.~Yamada, T.~Yagyu,
  Y.~Elovici, and A.~Shabtai, ``Deriving a cost-effective digital twin of an
  ics to facilitate security evaluation,'' in \emph{European Symposium on
  Research in Computer Security}.\hskip 1em plus 0.5em minus 0.4em\relax
  Springer, 2018, pp. 533--554.

\bibitem{lou2019idea}
X.~Lou, Y.~Guo, Y.~Gao, K.~Waedt, and M.~Parekh, ``An idea of using digital
  twin to perform the functional safety and cybersecurity analysis,'' in
  \emph{INFORMATIK 2019: 50 Jahre Gesellschaft f{\"u}r Informatik--Informatik
  f{\"u}r Gesellschaft (Workshop-Beitr{\"a}ge)}.\hskip 1em plus 0.5em minus
  0.4em\relax Gesellschaft f{\"u}r Informatik eV, 2019.

\bibitem{balta2019digital}
E.~C. Balta, D.~M. Tilbury, and K.~Barton, ``A digital twin framework for
  performance monitoring and anomaly detection in fused deposition modeling,''
  in \emph{2019 IEEE 15th International Conference on Automation Science and
  Engineering (CASE)}.\hskip 1em plus 0.5em minus 0.4em\relax IEEE, 2019, pp.
  823--829.

\bibitem{eckhart2018DT}
\BIBentryALTinterwordspacing
M.~Eckhart and A.~Ekelhart, ``A specification-based state replication approach
  for digital twins,'' ser. CPS-SPC '18.\hskip 1em plus 0.5em minus 0.4em\relax
  New York, NY, USA: Association for Computing Machinery, 2018, p. 36–47.
  [Online]. Available: \url{https://doi.org/10.1145/3264888.3264892}
\BIBentrySTDinterwordspacing

\bibitem{saad2020implementation}
A.~Saad, S.~Faddel, T.~Youssef, and O.~A. Mohammed, ``On the implementation of
  iot-based digital twin for networked microgrids resiliency against cyber
  attacks,'' \emph{IEEE transactions on smart grid}, vol.~11, no.~6, pp.
  5138--5150, 2020.

\bibitem{li2017dynamic}
C.~Li, S.~Mahadevan, Y.~Ling, S.~Choze, and L.~Wang, ``Dynamic bayesian network
  for aircraft wing health monitoring digital twin,'' \emph{Aiaa Journal},
  vol.~55, no.~3, pp. 930--941, 2017.

\bibitem{sleuters2019digital}
J.~Sleuters, Y.~Li, J.~Verriet, M.~Velikova, and R.~Doornbos, ``A digital twin
  method for automated behavior analysis of large-scale distributed iot
  systems,'' in \emph{2019 14th Annual Conference System of Systems Engineering
  (SoSE)}.\hskip 1em plus 0.5em minus 0.4em\relax IEEE, 2019, pp. 7--12.

\bibitem{wang2021digital}
H.~Wang, S.~Chen, M.~S. U.~I. Sami, F.~Rahman, and M.~Tehranipoor, ``Digital
  twin with a perspective from manufacturing industry,'' \emph{Emerging Topics
  in Hardware Security}, pp. 27--59, 2021.

\bibitem{jain2019digital}
P.~Jain, J.~Poon, J.~P. Singh, C.~Spanos, S.~R. Sanders, and S.~K. Panda, ``A
  digital twin approach for fault diagnosis in distributed photovoltaic
  systems,'' \emph{IEEE Transactions on Power Electronics}, vol.~35, no.~1, pp.
  940--956, 2019.

\bibitem{xu2019digital}
Y.~Xu, Y.~Sun, X.~Liu, and Y.~Zheng, ``A digital-twin-assisted fault diagnosis
  using deep transfer learning,'' \emph{IEEE Access}, vol.~7, pp.
  19\,990--19\,999, 2019.

\bibitem{kaewunruen2019digital}
S.~Kaewunruen and Q.~Lian, ``Digital twin aided sustainability-based lifecycle
  management for railway turnout systems,'' \emph{Journal of Cleaner
  Production}, vol. 228, pp. 1537--1551, 2019.

\bibitem{HIR2021}
\BIBentryALTinterwordspacing
``Reliability,'' in \emph{Heterogeneous Integration Roadmap}, P.~Wesling,
  Ed.\hskip 1em plus 0.5em minus 0.4em\relax IEEE Electronics Packaging
  Society, 2021, ch.~24, pp. 1--30. [Online]. Available:
  \url{https://eps.ieee.org/images/files/HIR\_2021/ch24\_rel.pdf}
\BIBentrySTDinterwordspacing

\bibitem{alves2019digital}
R.~G. Alves, G.~Souza, R.~F. Maia, A.~L.~H. Tran, C.~Kamienski, J.-P. Soininen,
  P.~T. Aquino, and F.~Lima, ``A digital twin for smart farming,'' in
  \emph{2019 IEEE Global Humanitarian Technology Conference (GHTC)}.\hskip 1em
  plus 0.5em minus 0.4em\relax IEEE, 2019, pp. 1--4.

\bibitem{tchana2019designing}
Y.~Tchana, G.~Ducellier, and S.~Remy, ``Designing a unique digital twin for
  linear infrastructures lifecycle management,'' \emph{Procedia CIRP}, vol.~84,
  pp. 545--549, 2019.

\bibitem{pokhrel2020digital}
A.~Pokhrel, V.~Katta, and R.~Colomo-Palacios, ``Digital twin for cybersecurity
  incident prediction: A multivocal literature review,'' in \emph{Proceedings
  of the IEEE/ACM 42nd International Conference on Software Engineering
  Workshops}, 2020, pp. 671--678.

\bibitem{rostami2014primer}
M.~Rostami, F.~Koushanfar, and R.~Karri, ``A primer on hardware security:
  Models, methods, and metrics,'' \emph{Proceedings of the IEEE}, vol. 102,
  no.~8, pp. 1283--1295, 2014.

\bibitem{asadizanjani2021physical}
N.~Asadizanjani, M.~T. Rahman, M.~Tehranipoor, and M.~H. Tehranipoor,
  \emph{Physical Assurance: For Electronic Devices and Systems}.\hskip 1em plus
  0.5em minus 0.4em\relax Springer, 2021.

\bibitem{rahman2018physical}
M.~T. Rahman, Q.~Shi, S.~Tajik, H.~Shen, D.~L. Woodard, M.~Tehranipoor, and
  N.~Asadizanjani, ``Physical inspection \& attacks: New frontier in hardware
  security,'' in \emph{2018 IEEE 3rd International Verification and Security
  Workshop (IVSW)}.\hskip 1em plus 0.5em minus 0.4em\relax IEEE, 2018, pp.
  93--102.

\bibitem{quadir2016survey}
S.~E. Quadir, J.~Chen, D.~Forte, N.~Asadizanjani, S.~Shahbazmohamadi, L.~Wang,
  J.~Chandy, and M.~Tehranipoor, ``A survey on chip to system reverse
  engineering,'' \emph{ACM journal on emerging technologies in computing
  systems (JETC)}, vol.~13, no.~1, pp. 1--34, 2016.

\bibitem{botero2021hardware}
U.~J. Botero, R.~Wilson, H.~Lu, M.~T. Rahman, M.~A. Mallaiyan, F.~Ganji,
  N.~Asadizanjani, M.~M. Tehranipoor, D.~L. Woodard, and D.~Forte, ``Hardware
  trust and assurance through reverse engineering: A tutorial and outlook from
  image analysis and machine learning perspectives,'' \emph{ACM Journal on
  Emerging Technologies in Computing Systems (JETC)}, vol.~17, no.~4, pp.
  1--53, 2021.

\bibitem{AMD}
\BIBentryALTinterwordspacing
M.~Sharma, ``Amd hardware security tricks can be bypassed with a shock of
  electricity,'' August 2021. [Online]. Available:
  \url{https://www.techradar.com/news/amd-hardware-security-tricks-can-be-bypassed-with-a-shock-of-electricity}
\BIBentrySTDinterwordspacing

\bibitem{TPM}
\BIBentryALTinterwordspacing
D.~Goodin, ``Trusted platform module security defeated in 30 minutes, no
  soldering required,'' August 2021. [Online]. Available:
  \url{https://arstechnica.com/gadgets/2021/08/how-to-go-from-stolen-pc-to-network-intrusion-in-30-minutes/}
\BIBentrySTDinterwordspacing

\bibitem{fournaris2017exploiting}
A.~P. Fournaris, L.~Pocero~Fraile, and O.~Koufopavlou, ``Exploiting hardware
  vulnerabilities to attack embedded system devices: A survey of potent
  microarchitectural attacks,'' \emph{Electronics}, vol.~6, no.~3, p.~52, 2017.

\bibitem{king2008designing}
S.~T. King, J.~Tucek, A.~Cozzie, C.~Grier, W.~Jiang, and Y.~Zhou, ``Designing
  and implementing malicious hardware.'' \emph{Leet}, vol.~8, pp. 1--8, 2008.

\bibitem{lee2020off}
D.~Lee, D.~Jung, I.~T. Fang, C.-C. Tsai, and R.~A. Popa, ``An off-chip attack
  on hardware enclaves via the memory bus,'' in \emph{29th $\{$USENIX$\}$
  Security Symposium ($\{$USENIX$\}$ Security 20)}, 2020.

\bibitem{grand2004practical}
J.~Grand, ``Practical secure hardware design for embedded systems,'' in
  \emph{Proceedings of the 2004 Embedded Systems Conference, San Francisco,
  California}, 2004.

\bibitem{rahman2021security}
M.~S. Rahman, A.~Nahiyan, F.~Rahman, S.~Fazzari, K.~Plaks, F.~Farahmandi,
  D.~Forte, and M.~Tehranipoor, ``Security assessment of dynamically obfuscated
  scan chain against oracle-guided attacks,'' \emph{ACM Transactions on Design
  Automation of Electronic Systems (TODAES)}, vol.~26, no.~4, pp. 1--27, 2021.

\bibitem{ahmed2021quantifiable}
B.~Ahmed, M.~K. Bepary, N.~Pundir, M.~Borza, O.~Raikhman, A.~Garg, D.~Donchin,
  A.~Cron, M.~A. Abdel-moneum, F.~Farahmandi \emph{et~al.}, ``Quantifiable
  assurance: From ips to platforms,'' \emph{Cryptology ePrint Archive}, 2021.

\bibitem{choo2020register}
H.~S. Choo, C.~Y. Ooi, M.~Inoue, N.~Ismail, M.~Moghbel, and C.~H. Kok,
  ``Register-transfer-level features for machine-learning-based hardware trojan
  detection,'' \emph{IEICE TRANSACTIONS on Fundamentals of Electronics,
  Communications and Computer Sciences}, vol. 103, no.~2, pp. 502--509, 2020.

\bibitem{he2019rtl}
M.~He, J.~Park, A.~Nahiyan, A.~Vassilev, Y.~Jin, and M.~Tehranipoor, ``Rtl-psc:
  Automated power side-channel leakage assessment at register-transfer level,''
  in \emph{2019 IEEE 37th VLSI Test Symposium (VTS)}.\hskip 1em plus 0.5em
  minus 0.4em\relax IEEE, 2019, pp. 1--6.

\bibitem{fallah2001occom}
F.~Fallah, S.~Devadas, and K.~Keutzer, ``Occom-efficient computation of
  observability-based code coverage metrics for functional verification,''
  \emph{IEEE Transactions on Computer-Aided Design of Integrated Circuits and
  Systems}, vol.~20, no.~8, pp. 1003--1015, 2001.

\bibitem{serrestou2007functional}
Y.~Serrestou, V.~Beroulle, and C.~Robach, ``Functional verification of rtl
  designs driven by mutation testing metrics,'' in \emph{10th Euromicro
  Conference on Digital System Design Architectures, Methods and Tools (DSD
  2007)}.\hskip 1em plus 0.5em minus 0.4em\relax IEEE, 2007, pp. 222--227.

\bibitem{goldstein1980scoap}
L.~H. Goldstein and E.~L. Thigpen, ``Scoap: Sandia
  controllability/observability analysis program,'' in \emph{Proceedings of the
  17th Design Automation Conference}, 1980, pp. 190--196.

\bibitem{waksman2013fanci}
A.~Waksman, M.~Suozzo, and S.~Sethumadhavan, ``Fanci: identification of
  stealthy malicious logic using boolean functional analysis,'' in
  \emph{Proceedings of the 2013 ACM SIGSAC conference on Computer \&
  communications security}, 2013, pp. 697--708.

\bibitem{chakraborty2009mero}
R.~S. Chakraborty, F.~Wolff, S.~Paul, C.~Papachristou, and S.~Bhunia, ``Mero: A
  statistical approach for hardware trojan detection,'' in \emph{International
  Workshop on Cryptographic Hardware and Embedded Systems}.\hskip 1em plus
  0.5em minus 0.4em\relax Springer, 2009, pp. 396--410.

\bibitem{cruz2018hardware}
J.~Cruz, F.~Farahmandi, A.~Ahmed, and P.~Mishra, ``Hardware trojan detection
  using atpg and model checking,'' in \emph{2018 31st international conference
  on VLSI design and 2018 17th international conference on embedded systems
  (VLSID)}.\hskip 1em plus 0.5em minus 0.4em\relax IEEE, 2018, pp. 91--96.

\bibitem{nahiyan2017code}
A.~Nahiyan and M.~Tehranipoor, ``Code coverage analysis for ip trust
  verification,'' in \emph{Hardware IP security and trust}.\hskip 1em plus
  0.5em minus 0.4em\relax Springer, 2017, pp. 53--72.

\bibitem{xu2017fast}
Q.~Xu and S.~Chen, ``Fast thermal analysis for fixed-outline 3d
  floorplanning,'' \emph{Integration}, vol.~59, pp. 157--167, 2017.

\bibitem{cong2004area}
J.~Cong, G.~Nataneli, M.~Romesis, and J.~R. Shinnerl, ``An area-optimality
  study of floorplanning,'' in \emph{Proceedings of the 2004 international
  symposium on Physical design}, 2004, pp. 78--83.

\bibitem{ma2020security}
H.~Ma, J.~He, Y.~Liu, L.~Liu, Y.~Zhao, and Y.~Jin, ``Security-driven placement
  and routing tools for electromagnetic side-channel protection,'' \emph{IEEE
  Transactions on Computer-Aided Design of Integrated Circuits and Systems},
  vol.~40, no.~6, pp. 1077--1089, 2020.

\bibitem{xiao2013bisa}
K.~Xiao and M.~Tehranipoor, ``Bisa: Built-in self-authentication for preventing
  hardware trojan insertion,'' in \emph{2013 IEEE international symposium on
  hardware-oriented security and trust (HOST)}.\hskip 1em plus 0.5em minus
  0.4em\relax IEEE, 2013, pp. 45--50.

\bibitem{salmani2009new}
H.~Salmani, M.~Tehranipoor, and J.~Plusquellic, ``New design strategy for
  improving hardware trojan detection and reducing trojan activation time,'' in
  \emph{2009 IEEE International Workshop on Hardware-Oriented Security and
  Trust}.\hskip 1em plus 0.5em minus 0.4em\relax IEEE, 2009, pp. 66--73.

\bibitem{vcd}
``Ieee standard for verilog hardware description language,'' \emph{IEEE Std
  1364-2005 (Revision of IEEE Std 1364-2001)}, pp. 325--348, 2006.

\bibitem{forte2017hardware}
D.~Forte, S.~Bhunia, and M.~M. Tehranipoor, \emph{Hardware protection through
  obfuscation}.\hskip 1em plus 0.5em minus 0.4em\relax Springer, 2017.

\bibitem{mack2008fundamental}
C.~Mack, \emph{Fundamental principles of optical lithography: the science of
  microfabrication}.\hskip 1em plus 0.5em minus 0.4em\relax John Wiley \& Sons,
  2008.

\bibitem{May2003fabrication}
G.~May and S.~Sze, \emph{Fundamentals of Semiconductor Fabrication},
  1st~ed.\hskip 1em plus 0.5em minus 0.4em\relax Wiley, 2003.

\bibitem{Campbell2008fabrication}
S.~Campbell, \emph{Fabrication Engineering at the Micro and Nanoscale},
  3rd~ed.\hskip 1em plus 0.5em minus 0.4em\relax Oxford University Press, 2008.

\bibitem{cobb2002hierarchical}
N.~B. Cobb and E.~Y. Sahouria, ``Hierarchical gdsii-based fracturing and job
  deck system,'' in \emph{21st Annual BACUS Symposium on Photomask Technology},
  vol. 4562.\hskip 1em plus 0.5em minus 0.4em\relax International Society for
  Optics and Photonics, 2002, pp. 734--742.

\bibitem{BACUS}
\BIBentryALTinterwordspacing
F.~E. Abboud, M.~Asturias, and M.~Chandramouli, ``Mask data processing in the
  era of multibeam writers,'' \emph{BACUS News}, Jan 2015. [Online]. Available:
  \url{https://spie.org/Documents/Membership/BacusNewsletters/BACUS-Newsletter-January-2015.pdf.}
\BIBentrySTDinterwordspacing

\bibitem{schulze2002gds}
S.~F. Schulze, P.~LaCour, and P.~D. Buck, ``Gds-based mask data preparation
  flow: data volume containment by hierarchical data processing,'' in
  \emph{22nd Annual BACUS Symposium on Photomask Technology}, vol. 4889.\hskip
  1em plus 0.5em minus 0.4em\relax International Society for Optics and
  Photonics, 2002, pp. 104--114.

\bibitem{hampden1995chemical}
M.~J. Hampden-Smith and T.~T. Kodas, ``Chemical vapor deposition of metals:
  Part 1. an overview of cvd processes,'' \emph{Chemical Vapor Deposition},
  vol.~1, no.~1, pp. 8--23, 1995.

\bibitem{smith1985ion}
W.~L. Smith, A.~Rosencwaig, and D.~L. Willenborg, ``Ion implant monitoring with
  thermal wave technology,'' \emph{Applied physics letters}, vol.~47, no.~6,
  pp. 584--586, 1985.

\bibitem{neubauer1996two}
G.~Neubauer, A.~Erickson, C.~C. Williams, J.~J. Kopanski, M.~Rodgers, and
  D.~Adderton, ``Two-dimensional scanning capacitance microscopy measurements
  of cross-sectioned very large scale integration test structures,''
  \emph{Journal of Vacuum Science \& Technology B: Microelectronics and
  Nanometer Structures Processing, Measurement, and Phenomena}, vol.~14, no.~1,
  pp. 426--432, 1996.

\bibitem{vandervorst2001towards}
W.~Vandervorst, P.~Eyben, S.~Callewaert, T.~Hantschel, N.~Duhayon, M.~Xu,
  T.~Trenkler, and T.~Clarysse, ``Towards routine, quantitative two-dimensional
  carrier profiling with scanning spreading resistance microscopy,'' in
  \emph{AIP Conference Proceedings}, vol. 550, no.~1.\hskip 1em plus 0.5em
  minus 0.4em\relax American Institute of Physics, 2001, pp. 613--619.

\bibitem{bushnell2004essentials}
M.~Bushnell and V.~Agrawal, \emph{Essentials of electronic testing for digital,
  memory and mixed-signal VLSI circuits}.\hskip 1em plus 0.5em minus
  0.4em\relax Springer Science \& Business Media, 2004, vol.~17.

\bibitem{andrea2017}
A.~Chen and R.~H.-Y. Lo, \emph{Semiconductor Packaging: Materials Interaction
  and Reliability}.\hskip 1em plus 0.5em minus 0.4em\relax CRC Press, 2017.

\bibitem{nourani2008low}
M.~Nourani, M.~Tehranipoor, and N.~Ahmed, ``Low-transition test pattern
  generation for bist-based applications,'' \emph{IEEE Transactions on
  Computers}, vol.~57, no.~3, pp. 303--315, 2008.

\bibitem{tehranipour2003testing}
M.~H. Tehranipour, N.~Ahmed, and M.~Nourani, ``Testing soc interconnects for
  signal integrity using boundary scan,'' in \emph{Proceedings. 21st VLSI Test
  Symposium, 2003.}\hskip 1em plus 0.5em minus 0.4em\relax IEEE, 2003, pp.
  158--163.

\bibitem{tehranipoor2010survey}
M.~Tehranipoor and F.~Koushanfar, ``A survey of hardware trojan taxonomy and
  detection,'' \emph{IEEE design \& test of computers}, vol.~27, no.~1, pp.
  10--25, 2010.

\bibitem{patil2017manufacturer}
V.~C. Patil, A.~Vijayakumar, and S.~Kundu, ``Manufacturer turned attacker:
  Dangers of stealthy trojans via threshold voltage manipulation,'' in
  \emph{2017 IEEE North Atlantic Test Workshop (NATW)}.\hskip 1em plus 0.5em
  minus 0.4em\relax IEEE, 2017, pp. 1--6.

\bibitem{contreras2013secure}
G.~K. Contreras, M.~T. Rahman, and M.~Tehranipoor, ``Secure split-test for
  preventing ic piracy by untrusted foundry and assembly,'' in \emph{2013 IEEE
  International symposium on defect and fault tolerance in VLSI and
  nanotechnology systems (DFTS)}.\hskip 1em plus 0.5em minus 0.4em\relax IEEE,
  2013, pp. 196--203.

\bibitem{huang2012parametric}
K.~Huang, J.~M. Carulli, and Y.~Makris, ``Parametric counterfeit ic detection
  via support vector machines,'' in \emph{2012 IEEE International Symposium on
  Defect and Fault Tolerance in VLSI and Nanotechnology Systems (DFT)}.\hskip
  1em plus 0.5em minus 0.4em\relax IEEE, 2012, pp. 7--12.

\bibitem{zhang2012path}
X.~Zhang, K.~Xiao, and M.~Tehranipoor, ``Path-delay fingerprinting for
  identification of recovered ics,'' in \emph{2012 IEEE International symposium
  on defect and fault tolerance in VLSI and nanotechnology systems
  (DFT)}.\hskip 1em plus 0.5em minus 0.4em\relax IEEE, 2012, pp. 13--18.

\bibitem{nahiyan2017hardware}
A.~Nahiyan, M.~Sadi, R.~Vittal, G.~Contreras, D.~Forte, and M.~Tehranipoor,
  ``Hardware trojan detection through information flow security verification,''
  in \emph{2017 IEEE International Test Conference (ITC)}.\hskip 1em plus 0.5em
  minus 0.4em\relax IEEE, 2017, pp. 1--10.

\bibitem{hu2014gate}
W.~Hu, D.~Mu, J.~Oberg, B.~Mao, M.~Tiwari, T.~Sherwood, and R.~Kastner,
  ``Gate-level information flow tracking for security lattices,'' \emph{ACM
  Transactions on Design Automation of Electronic Systems (TODAES)}, vol.~20,
  no.~1, pp. 1--25, 2014.

\bibitem{liu2016silicon}
Y.~Liu, Y.~Jin, A.~Nosratinia, and Y.~Makris, ``Silicon demonstration of
  hardware trojan design and detection in wireless cryptographic ics,''
  \emph{IEEE Transactions on Very Large Scale Integration (VLSI) Systems},
  vol.~25, no.~4, pp. 1506--1519, 2016.

\bibitem{kison2019security}
C.~Kison, O.~M. Awad, M.~Fyrbiak, and C.~Paar, ``Security implications of
  intentional capacitive crosstalk,'' \emph{IEEE Transactions on Information
  Forensics and Security}, vol.~14, no.~12, pp. 3246--3258, 2019.

\bibitem{rizvi2018handbook}
S.~Rizvi, \emph{Handbook of photomask manufacturing technology}.\hskip 1em plus
  0.5em minus 0.4em\relax CRC Press, 2018.

\bibitem{takahashi2000proximity}
K.~Takahashi, M.~Osawa, M.~Sato, H.~Arimoto, K.~Ogino, H.~Hoshino, and
  Y.~Machida, ``Proximity effect correction using pattern shape modification
  and area density map,'' \emph{Journal of Vacuum Science \& Technology B:
  Microelectronics and Nanometer Structures Processing, Measurement, and
  Phenomena}, vol.~18, no.~6, pp. 3150--3157, 2000.

\bibitem{hasegawa2017trojan}
K.~Hasegawa, M.~Yanagisawa, and N.~Togawa, ``Trojan-feature extraction at
  gate-level netlists and its application to hardware-trojan detection using
  random forest classifier,'' in \emph{2017 IEEE International Symposium on
  Circuits and Systems (ISCAS)}.\hskip 1em plus 0.5em minus 0.4em\relax IEEE,
  2017, pp. 1--4.

\bibitem{STDF}
\BIBentryALTinterwordspacing
M.~Sharma, ``Standard test data format specification,'' August 2021. [Online].
  Available:
  \url{http://www.kanwoda.com/wp-content/uploads/2015/05/std-spec.pdf}
\BIBentrySTDinterwordspacing

\bibitem{guin2013anti}
U.~Guin, D.~Forte, and M.~Tehranipoor, ``Anti-counterfeit techniques: From
  design to resign,'' in \emph{2013 14th International workshop on
  microprocessor test and verification}.\hskip 1em plus 0.5em minus 0.4em\relax
  IEEE, 2013, pp. 89--94.

\bibitem{koller2007introduction}
D.~Koller, N.~Friedman, S.~D{\v{z}}eroski, C.~Sutton, A.~McCallum, A.~Pfeffer,
  P.~Abbeel, M.-F. Wong, C.~Meek, J.~Neville \emph{et~al.}, \emph{Introduction
  to statistical relational learning}.\hskip 1em plus 0.5em minus 0.4em\relax
  MIT press, 2007.

\bibitem{guo2002survey}
H.~Guo and W.~Hsu, ``A survey of algorithms for real-time bayesian network
  inference,'' in \emph{Join Workshop on Real Time Decision Support and
  Diagnosis Systems}, 2002.

\bibitem{chen2012good}
S.~H. Chen and C.~A. Pollino, ``Good practice in bayesian network modelling,''
  \emph{Environmental Modelling \& Software}, vol.~37, pp. 134--145, 2012.

\bibitem{heckerman1995learning}
D.~Heckerman, D.~Geiger, and D.~M. Chickering, ``Learning bayesian networks:
  The combination of knowledge and statistical data,'' \emph{Machine learning},
  vol.~20, no.~3, pp. 197--243, 1995.

\bibitem{rabiner1989tutorial}
L.~R. Rabiner, ``A tutorial on hidden markov models and selected applications
  in speech recognition,'' \emph{Proceedings of the IEEE}, vol.~77, no.~2, pp.
  257--286, 1989.

\bibitem{rabiner1986introduction}
L.~Rabiner and B.~Juang, ``An introduction to hidden markov models,''
  \emph{ieee assp magazine}, vol.~3, no.~1, pp. 4--16, 1986.

\bibitem{forney1973viterbi}
G.~D. Forney, ``The viterbi algorithm,'' \emph{Proceedings of the IEEE},
  vol.~61, no.~3, pp. 268--278, 1973.

\bibitem{richardson2006markov}
M.~Richardson and P.~Domingos, ``Markov logic networks,'' \emph{Machine
  learning}, vol.~62, no. 1-2, pp. 107--136, 2006.

\bibitem{lowd2007efficient}
D.~Lowd and P.~Domingos, ``Efficient weight learning for markov logic
  networks,'' in \emph{European conference on principles of data mining and
  knowledge discovery}.\hskip 1em plus 0.5em minus 0.4em\relax Springer, 2007,
  pp. 200--211.

\bibitem{skorobogatov2012breakthrough}
S.~Skorobogatov and C.~Woods, ``Breakthrough silicon scanning discovers
  backdoor in military chip,'' in \emph{International Workshop on Cryptographic
  Hardware and Embedded Systems}.\hskip 1em plus 0.5em minus 0.4em\relax
  Springer, 2012, pp. 23--40.

\bibitem{laptev2015generic}
N.~Laptev, S.~Amizadeh, and I.~Flint, ``Generic and scalable framework for
  automated time-series anomaly detection,'' in \emph{Proceedings of the 21th
  ACM SIGKDD international conference on knowledge discovery and data mining},
  2015, pp. 1939--1947.

\bibitem{shahbazmohamadi2014advanced}
S.~Shahbazmohamadi, D.~Forte, and M.~Tehranipoor, ``Advanced physical
  inspection methods for counterfeit ic detection,'' in \emph{ISTFA 2014:
  Conference Proceedings from the 40th International Symposium for Testing and
  Failure Analysis}.\hskip 1em plus 0.5em minus 0.4em\relax ASM International,
  2014, p.~55.

\bibitem{baygin2017machine}
M.~Baygin, M.~Karakose, A.~Sarimaden, and A.~Erhan, ``Machine vision based
  defect detection approach using image processing,'' in \emph{2017
  international artificial intelligence and data processing symposium
  (IDAP)}.\hskip 1em plus 0.5em minus 0.4em\relax Ieee, 2017, pp. 1--5.

\bibitem{ghosh2018recycled}
P.~Ghosh and R.~S. Chakraborty, ``Recycled and remarked counterfeit integrated
  circuit detection by image-processing-based package texture and indent
  analysis,'' \emph{IEEE Transactions on Industrial Informatics}, vol.~15,
  no.~4, pp. 1966--1974, 2018.

\bibitem{shakya2017benchmarking}
B.~Shakya, T.~He, H.~Salmani, D.~Forte, S.~Bhunia, and M.~Tehranipoor,
  ``Benchmarking of hardware trojans and maliciously affected circuits,''
  \emph{Journal of Hardware and Systems Security}, vol.~1, no.~1, pp. 85--102,
  2017.

\bibitem{salmani2013design}
H.~Salmani, M.~Tehranipoor, and R.~Karri, ``On design vulnerability analysis
  and trust benchmarks development,'' in \emph{2013 IEEE 31st international
  conference on computer design (ICCD)}.\hskip 1em plus 0.5em minus 0.4em\relax
  IEEE, 2013, pp. 471--474.

\bibitem{meng2021rtl}
X.~Meng, S.~Kundu, A.~K. Kanuparthi, and K.~Basu, ``Rtl-contest: Concolic
  testing on rtl for detecting security vulnerabilities,'' \emph{IEEE
  Transactions on Computer-Aided Design of Integrated Circuits and Systems},
  2021.

\bibitem{ardeshiricham2017register}
A.~Ardeshiricham, W.~Hu, J.~Marxen, and R.~Kastner, ``Register transfer level
  information flow tracking for provably secure hardware design,'' in
  \emph{Design, Automation \& Test in Europe Conference \& Exhibition (DATE),
  2017}.\hskip 1em plus 0.5em minus 0.4em\relax IEEE, 2017, pp. 1691--1696.

\bibitem{vashistha2018detecting}
N.~Vashistha, M.~T. Rahman, H.~Shen, D.~L. Woodard, N.~Asadizanjani, and
  M.~Tehranipoor, ``Detecting hardware trojans inserted by untrusted foundry
  using physical inspection and advanced image processing,'' \emph{Journal of
  Hardware and Systems Security}, vol.~2, no.~4, pp. 333--344, 2018.

\bibitem{marcot2006guidelines}
B.~G. Marcot, J.~D. Steventon, G.~D. Sutherland, and R.~K. McCann, ``Guidelines
  for developing and updating bayesian belief networks applied to ecological
  modeling and conservation,'' \emph{Canadian Journal of Forest Research},
  vol.~36, no.~12, pp. 3063--3074, 2006.

\bibitem{beedkar2013fully}
K.~Beedkar, L.~D. Corro, and R.~Gemulla, ``Fully parallel inference in markov
  logic networks,'' \emph{Datenbanksysteme f{\"u}r Business, Technologie und
  Web (BTW) 2026}, 2013.

\bibitem{shavlik2009speeding}
J.~Shavlik and S.~Natarajan, ``Speeding up inference in markov logic networks
  by preprocessing to reduce the size of the resulting grounded network,'' in
  \emph{Twenty-First International Joint Conference on Artificial
  Intelligence}, 2009.

\bibitem{daly2009learning}
R.~Daly and Q.~Shen, ``Learning bayesian network equivalence classes with ant
  colony optimization,'' \emph{Journal of Artificial Intelligence Research},
  vol.~35, pp. 391--447, 2009.

\bibitem{nath2018system}
A.~P.~D. Nath, S.~Ray, A.~Basak, and S.~Bhunia, ``System-on-chip security
  architecture and cad framework for hardware patch,'' in \emph{2018 23rd Asia
  and South Pacific Design Automation Conference (ASP-DAC)}.\hskip 1em plus
  0.5em minus 0.4em\relax IEEE, 2018, pp. 733--738.

\bibitem{beyrouthy2007novel}
T.~Beyrouthy, A.~Razafindraibe, L.~Fesquet, M.~Renaudin, S.~Chaudhuri,
  S.~Guilley, P.~Hoogvorst, and J.-L. Danger, ``A novel asynchronous e-fpga
  architecture for security applications,'' in \emph{2007 International
  Conference on Field-Programmable Technology}.\hskip 1em plus 0.5em minus
  0.4em\relax IEEE, 2007, pp. 369--372.

\bibitem{saha2020fpga}
S.~K. Saha and C.~Bobda, ``Fpga accelerated embedded system security through
  hardware isolation,'' in \emph{2020 Asian Hardware Oriented Security and
  Trust Symposium (AsianHOST)}.\hskip 1em plus 0.5em minus 0.4em\relax IEEE,
  2020, pp. 1--6.

\bibitem{delshadtehrani2020phmon}
L.~Delshadtehrani, S.~Canakci, B.~Zhou, S.~Eldridge, A.~Joshi, and M.~Egele,
  ``Phmon: a programmable hardware monitor and its security use cases,'' in
  \emph{29th $\{$USENIX$\}$ Security Symposium ($\{$USENIX$\}$ Security 20)},
  2020, pp. 807--824.

\end{thebibliography}

\end{document}